\documentclass[intlimits,twoside,a4paper]{article}

\usepackage[cp1251]{inputenc}

\usepackage[eqsecnum]{cmpj3}

\usepackage{bm}

\articletype{Review}

\issue{2021}{24}{4}{42701}
\doinumber{10.5488/CMP.24.42701}

\title[Itinerant ferromagnetism in narrow-band metals]
{Itinerant ferromagnetism in narrow-band metals}

\author[P. Farka\v sovsk\'y]{P. Farka\v sovsk\'y}
\address{
Institute  of  Experimental  Physics,  Slovak   Academy   of
Sciences,\\
Watsonova 47, 043 53 Ko\v {s}ice, Slovakia}


\date{Received August 11, 2021, in final form October 19, 2021}
\begin{document}

\maketitle

\begin{abstract}
Since its introduction in 1963, the Hubbard model has becomes one of the 
most popular models used in the literature to study cooperative phenomena 
in narrow-band metals (ferromagnetism, metal-insulator transitions,
charge-density waves, high-T$_c$ superconductivity).  
Amongst all these cooperative phenomena, the problem of 
itinerant ferromagnetism in the Hubbard model has the longest history.
However, in spite of an impressive research activity in the past, the 
underlying physics (microscopic mechanisms) that leads to the stabilization 
of itinerant ferromagnetism in Hubbard model (narrow-band metals) is still 
far from being understood. In this review we present our numerical results
concerning this subject, which have been reached by small cluster exact 
diagonalization, density matrix renormalization group  and quantum Monte Carlo 
calculations within various extensions of the Hubbard model. Particular attention 
is paid to a description of crucial mechanisms (interactions) that support the 
stabilization of the ferromagnetic state, and namely: (i) the long-range 
hopping, (ii) the correlated hopping, (iii) the long-range Coulomb interaction, 
(iv) the flat bands and (v) the lattice structure. Most of the presented results 
have been obtained for the one-dimensional case, but the influence of the increasing
dimension of the system on the ferromagnetic state is also intensively discussed. 
\printkeywords
\end{abstract}


\section{Introduction}
The microscopic description of itinerant ferromagnetism in narrow-band
metals is one of the most interesting as well as the most complicated many-particle 
problems in condensed matter physics. This is expected to be due to the 
interplay between ordinary, spin-independent Coulomb interaction 
(strong and strongly screened) and kinetic energy of itinerant 
electrons within the frames determined by the Pauli exclusion principle. 
The single-band Hubbard model~\cite{Hubbard},
possibly the simplest lattice model of correlated electrons, was first
thought to encompass a minimal description of band-ferromagnetism. 
The Hamiltonian of this model can be written as a sum of two terms:
\begin{equation}
H=\sum_{ij\sigma}t_{ij}c^+_{i\sigma}c_{j\sigma}+
U\sum_{i}n_{i\uparrow}n_{i\downarrow},
\label{Eq1}
\end{equation}
where $c^+_{i\sigma}$ and $c_{i\sigma}$ are the creation and annihilation
operators  for an electron of spin $\sigma=\uparrow,\downarrow$ at site $i$ and 
$n_{i\sigma}$ is the corresponding number operator, which counts
the number of electrons of spin $\sigma$ on site $i$.

The first term of (\ref{Eq1}) is the kinetic energy of itinerant electrons. 
It corresponds to quantum-mechanical hopping of the $\sigma$-spin electrons 
between sites $i$ and $j$, with transition hopping probabilities $t_{ij}$. Usually, 
it is assumed (the ordinary Hubbard model) that $t_{ij}=-t$ if $i$ and $j$
are nearest neighbors and $t_{ij}=0$ otherwise.

The second term corresponds to the Coulomb repulsion between two electrons
of opposite spins at the same site. The long-range contribution is assumed
to be screened and only the interaction when both the $\sigma=\uparrow$ 
and $\sigma=\downarrow$ electrons are on the same atom is retained,
yielding an additional energy of $U$ when the atom is doubly occupied.

Thus, naively the ordinary Hubbard model might be thought to describe
the competition between the kinetic energy and the short-range Coulomb
interaction, but in fact there is the third ``force'' at work which severely
complicates the problem: Pauli exclusion. Electrons are fermions and so   
the many-particle wave function must be antisymmetric under interchange 
of any two electrons. Unlike the first two forces which are basically
short-range interactions, antisymmetrizing wave functions are effectively 
a long-range interaction that significantly complicates a description of
correlation effects within, at the first glance, the very simple model.   
On the other hand, there were very pleasurable arguments which, at least
in the first stages of the study, favored the Hubbard model as  the generic 
model for a description of the band ferrommagnetism in the transition 
metal compounds. Indeed, writing the magnetization as 
$m_i=n_{i\uparrow}-n_{i\downarrow}$ and total number of $n$ electrons (per
site) as $n_i=n_{i\uparrow}+n_{i\downarrow}$, the interaction part of the Hubbard
model can be rewritten as 
$Un_{i\uparrow}n_{i\downarrow}=U(n_i^2-m_i^2)/4$. Since the number $n$ of
electron per atom is fixed, the Coulomb interaction favors the formation
of a magnetic moment. 

However, the subsequent studies of the model 
showed  that the single-band Hubbard model is not the
canonical model for ferromagnetism. Indeed, the existence of saturated
ferromagnetism has been proven rigorously only for very special limits.
The first well-known example is the Nagaoka ferromagnetism that
comes from the Hubbard model in the limit of infinite repulsion and
one hole in a half-filled band~\cite{Nagaoka}.
Another example, where saturated ferromagnetism has been shown to
exist, is the case of the one-dimensional Hubbard model with
nearest and next-nearest-neighbor hopping at low electron
densities~\cite{M_H}.
Furthermore, several examples of the fully polarized ground state 
have been found on special lattices as are the bipartite lattices 
with sublattices containing a different number of sites~\cite{Lieb},
the fcc-type lattices~\cite{Ulmke,Pandey},
the lattices with long-range electron 
hopping~\cite{Pieri, Pieri_a,Pieri_b},
the flat bands~\cite{FB2,FB3,FB4,FB4a,FB5} and the nearly flat-band
systems~\cite{Khodel,Volovik1,Gofron,NFB1,NFB2,NFB3,NFB4,Gulacsi,Volovik2}.
This indicates that the lattice structure, which dictates the shape
of the density of states (DOS), plays an important role in stabilizing 
the ferromagnetic state.  

To examine in more detail the role of these (and some other) factors 
on the stabilization of the ferromagnetic state in the generalized 
Hubbard model, in our previous studies we have used  
small-cluster-exact diagonalization~\cite{Lanczos}, 
density matrix renormalization group~\cite{Peschel,Schollwock,Hallberg} 
and projector-quantum Monte Carlo\cite{Sorella,Loh,Imada}  calculations. 
In particular, we have examined 
the role of the long-range hopping~\cite{Fark1,Fark2} and 
long-range Coulomb interaction~\cite{Fark3} with exponentially decaying amplitudes,
the correlated hopping~\cite{Fark4}, the flat bands~\cite{Fark5} 
and the lattice structure~\cite{Fark6,Fark7}.
The main results of our numerical studies are summarized in this 
short review.

\section{Results and discussion}
\subsection{The effect of long-range hopping}
\subsubsection{One-dimensional case}

Since the model including the electron hopping only  to the nearest neighbors 
may seem at first glance a very crude approximation, in order to have a more 
realistic description of electron processes in transition metal compounds, 
we have  generalized this model  by taking into account also transitions 
to next neighbors. Basically, there are two possible ways of performing such 
a generalization. The first way is to assign independent transition amplitudes 
for the first  ($t_1$), second ($t_2$), third ($t_3$), forth ($t_4$),... nearest 
neighbors, while the second way is to describe the electron hopping by a simple 
one-parametric formula~\cite{Fark8,Fark9} with exponentially decaying hopping 
amplitudes between $\bf R_i$ and $\bf R_j$ lattice sites, i.e., 
\begin{equation}
t_{i,j}(q)=\left \{ \begin{array}{ll}
  \quad 0,               &   {\bf R_i}={\bf R_j},\\
  -q^{|{\bf R_i}-{\bf R_j}|}/q,      &   {\bf R_i}\neq {\bf R_j},
\end{array}
\right.
\label{Eq2}
\end{equation}
where $q$ is the parameter that controls the range of electron hopping
($0\leqslant q \leqslant 1$). 
From the practical point of view, the second method is more suitable because 
it does not expand the model parameter space and has a clearer physical meaning, 
since the  atomic wave functions have also the exponential decay with increasing 
distance. For this reason,  for a  description of electron hopping in
the generalized model, we have chosen the long-range hopping with exponentially decreasing
amplitudes.

\begin{figure}[!t]
\begin{center}
\includegraphics[width=7cm]{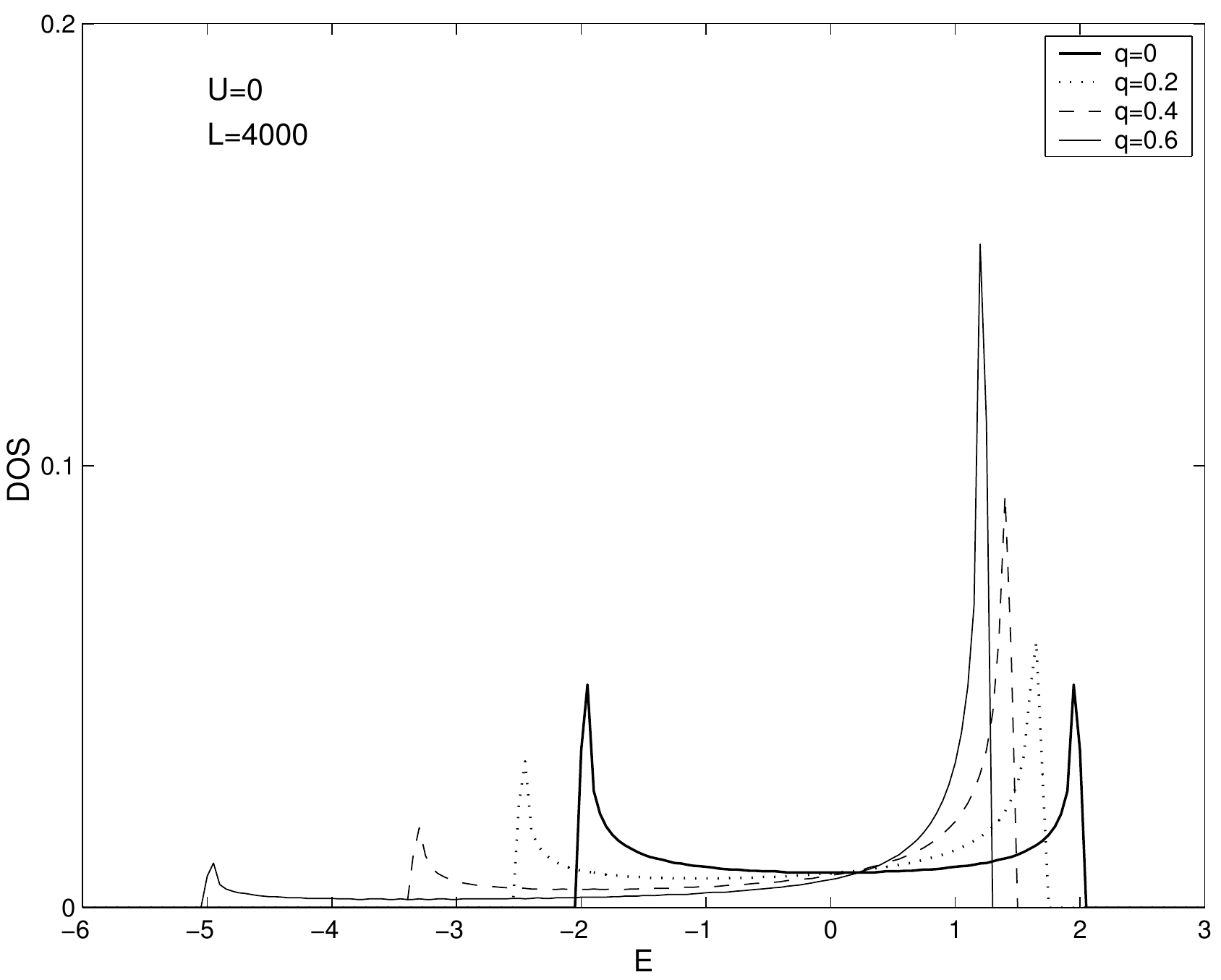}
\end{center}
\caption{ The non-interacting ($D=1$) DOS corresponding to the long-range
hopping for different $q$ and $L=4000$~\cite{Fark1}.}
\label{fig1}
\end{figure}

The selection of hopping matrix elements in the form given by equation~(\ref{Eq2}) 
has several advantages. It represents a much more realistic
type of electron hopping on a lattice (in comparison to
nearest-neighbor hopping), and it allows us to change
continuously the type of hopping (band) from nearest-neighbor $(q=0)$
to  infinite-range $(q=1)$ hopping and thus immediately study the effect 
of the long-range hopping. Another advantage follows from figure~\ref{fig1}, 
where the density of states (DOS) corresponding to equation~(\ref{Eq2}) is displayed
for several values of $q$. 
It is seen that with increasing $q$, more weight
shifts to the upper edge of the band and the DOS becomes strongly asymmetric.
Thus, one can simultaneously study  (by changing only one parameter $q$)
the influence of the increasing  asymmetry in the DOS
and the influence of the long-range hopping on the ground state properties of
the Hubbard model.

The Hamiltonian of the single-band Hubbard model with long-range hopping is given
by:
\begin{equation}
H=\sum_{ij\sigma}t_{ij}(q)c^+_{i\sigma}c_{j\sigma}+
U\sum_{i}n_{i\uparrow}n_{i\downarrow}.
\label{Eq3}
\end{equation}
The exact results on the ground states of the Hubbard model with 
the generalized type of hopping (\ref{Eq3}) exist only for the special 
case of $q=1$ when the electrons can hop to all sites with equal 
probabilities~\cite{Pieri, Pieri_a, Pieri_b}. For this type of hopping and the electron 
filling just above half-filling ($N=\sum_{\sigma}N_{\sigma}=L+1$,
where $L$ is the number of lattice sites), it was shown that the ground 
state is not degenerate with respect to the total spin $S$ and it is 
maximum ferromagnetic with $S=(L-1)/2$ (for all $U>0$). For higher 
fillings $(N > L+1)$, the ferromagnetic ground state still exists 
but it is completely degenerate with respect to $S$. The limit of 
infinite-range hopping is, however, the least realistic limit of
equation~(\ref{Eq2}). 
It is interesting, therefore, to look at the possibility of ferromagnetism 
in the Hubbard model with a generalized type of hopping for smaller values 
of $q$ that describe a much more realistic type of electron hopping.

In our paper~\cite{Fark1} we have extended calculations to arbitrary $q$ and  
arbitrary band fillings $n=N/L$. The ground states of the model 
have been determined by exact diagonalizations for a wide range of model
parameters ($q,U,n$). Typical examples are  chosen from a large 
number of available results to represent the most interesting cases.
The results obtained are presented in the form of phase diagrams 
in the $U$--$q$ plane. To determine the phase diagram in the $U$--$q$
plane (corresponding to some $L$ and $n$), the ground state energy 
of the model is calculated point by point as functions of $q$ and $U$. 
Of course, such a procedure demands in practice a considerable amount 
of CPU time, which imposes severe restrictions upon the size of clusters 
that can be studied using this method~($L \sim 16$).
Fortunately, we have found that the ground-state energy of the 
model depends on $L$ only very weakly (for a  wide  range of the 
model parameters) and thus already such small clusters can satisfactorily describe 
 the ground state properties of the model.

\begin{figure}[!b]
\begin{center}
\includegraphics[width=12cm]{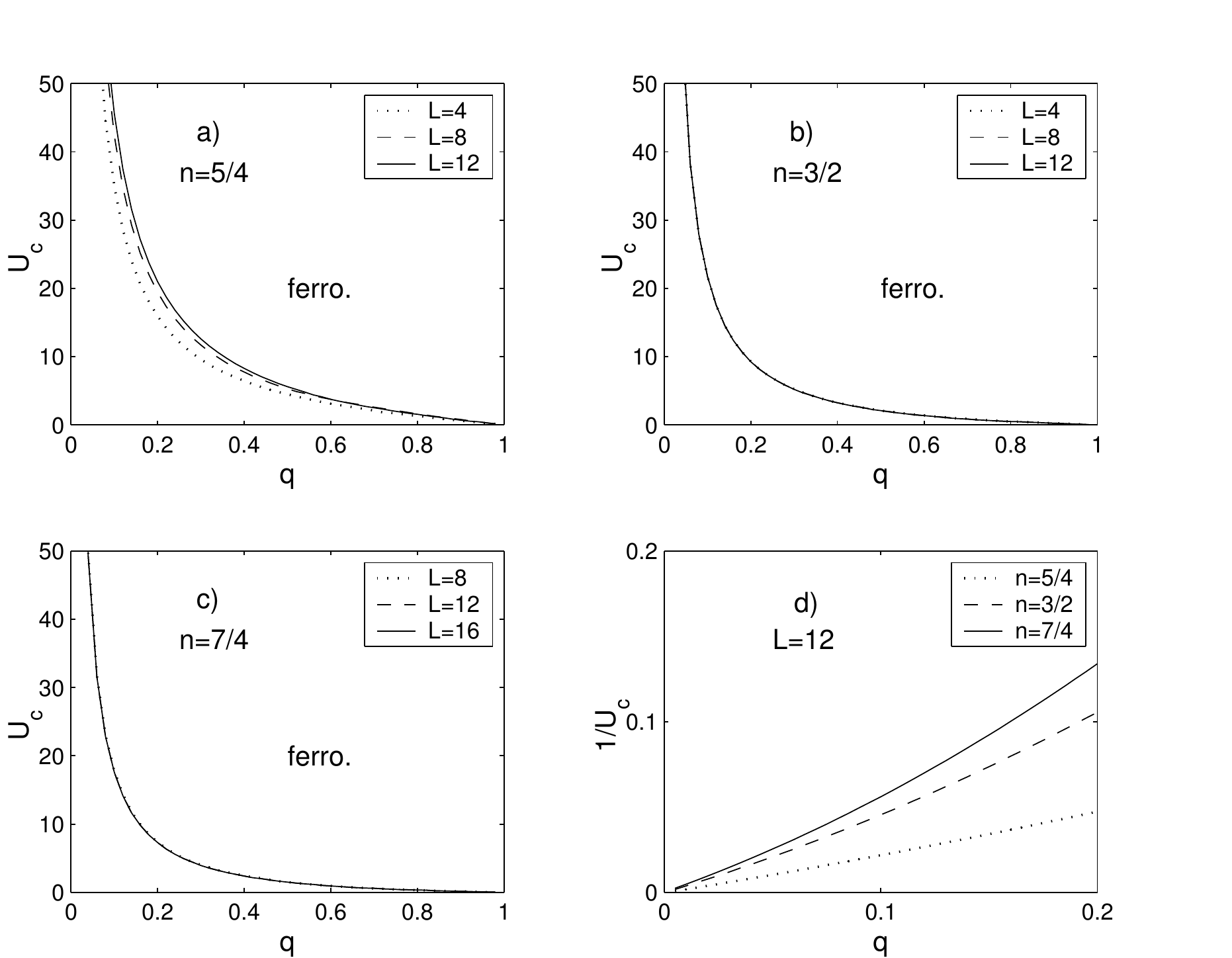}
\end{center}
\caption{ The critical interaction strength $U_{c}$ ($1/U_c$) as a function 
of $q$ calculated for different $n$ and~$L$~\cite{Fark1}.}
\label{fig2}
\end{figure}

Although the appearance of the ferromagnetic state at  $q=1$ and $N=L+1$ 
(discussed above) is interesting from the theoretical point of view, 
in the thermodynamic limit ($L \to \infty$) this result is not significant if the ferromagnetic 
state does not persist also for higher fillings. Analytical results 
obtained for $q=1$ predict, however, that the ground states for 
$N > L+1$ are completely degenerate with respect to the total spin 
$S$ and thus the only possibility for the stabilization of the 
ferromagnetic state is that the long-range hopping with $q\neq 1$ 
removes this degeneracy. Numerical calculations that we have performed 
for a wide range of electron
fillings $n>1$ fully confirmed this assumption. It was found that 
the long-range hopping with $q\neq 1$ not only removes the degeneracy 
of the ground states with respect to $S$ but at the same time stabilizes 
the ferromagnetic state. Furthermore, these calculations showed that 
the effect of the long-range hopping on the stability of the ferromagnetic 
state is extremely strong, especially for small values of $q$.  
The results of our small-cluster exact-diagonalization calculations
obtained on finite clusters up to $L=16$ sites are summarized in 
figure~\ref{fig2}. 
There is shown a critical interaction strength $U_c$, above 
which the ground state is ferromagnetic, as a function of $q$ for several 
values of electron concentrations $n$ ($n=5/4\,,3/2\,,7/4$). 
To reveal the finite-size effects on the stability of ferromagnetic domains, 
the behavior of the critical interaction strength $U_c(q)$ has been calculated
on several finite clusters at each electron filling. It is seen that 
finite-size effects on $U_c$ are small and thus these results can be 
satisfactorily extrapolated to the thermodynamic limit $L\to \infty$.           
Our results clearly demonstrate that the ferromagnetic state is
strongly influenced by $q$ for electron concentrations above half-filling
and generally it is stabilized with increasing $q$. The effect 
is especially strong for small values of $q$ where small changes of 
$q$ reduce dramatically the critical interaction strength $U_c$ and
so the ferromagnetic state becomes stable for a wide range of model 
parameters.  The results presented in figure~\ref{fig2}d show that only for 
$q=0$ (nearest-neighbor hopping) $U_c=\infty$, while for finite 
$q$ (that represents a much more realistic type of electron 
hopping), the critical interaction strength $U_c$ is finite. Thus, the
absence of ferromagnetism in the ordinary Hubbard model with the
nearest-neighbor hopping ($q=0$) can be explained as a consequence of 
too simplified description of electron hopping on the lattice. For 
any $q>0$, ferromagnetism comes naturally from the Hubbard model with 
a generalized type of hopping for a wide range of model parameters 
without any other assumptions.  This opens up a new  route towards 
the understanding of ferromagnetism in the Hubbard model.

\subsubsection{Two-dimensional case}

 We have also performed  the same calculations for the physically more
interesting two-dimensional case~\cite{Fark2}. Since in the two-dimensional case only
very small clusters ($L \sim 20$) are accessible by the exact diagonalization
method, we support these calculations by the quantum Monte Carlo method
that can treat several times larger clusters. 

\begin{figure}[!t]
\begin{center}
\includegraphics[width=10cm]{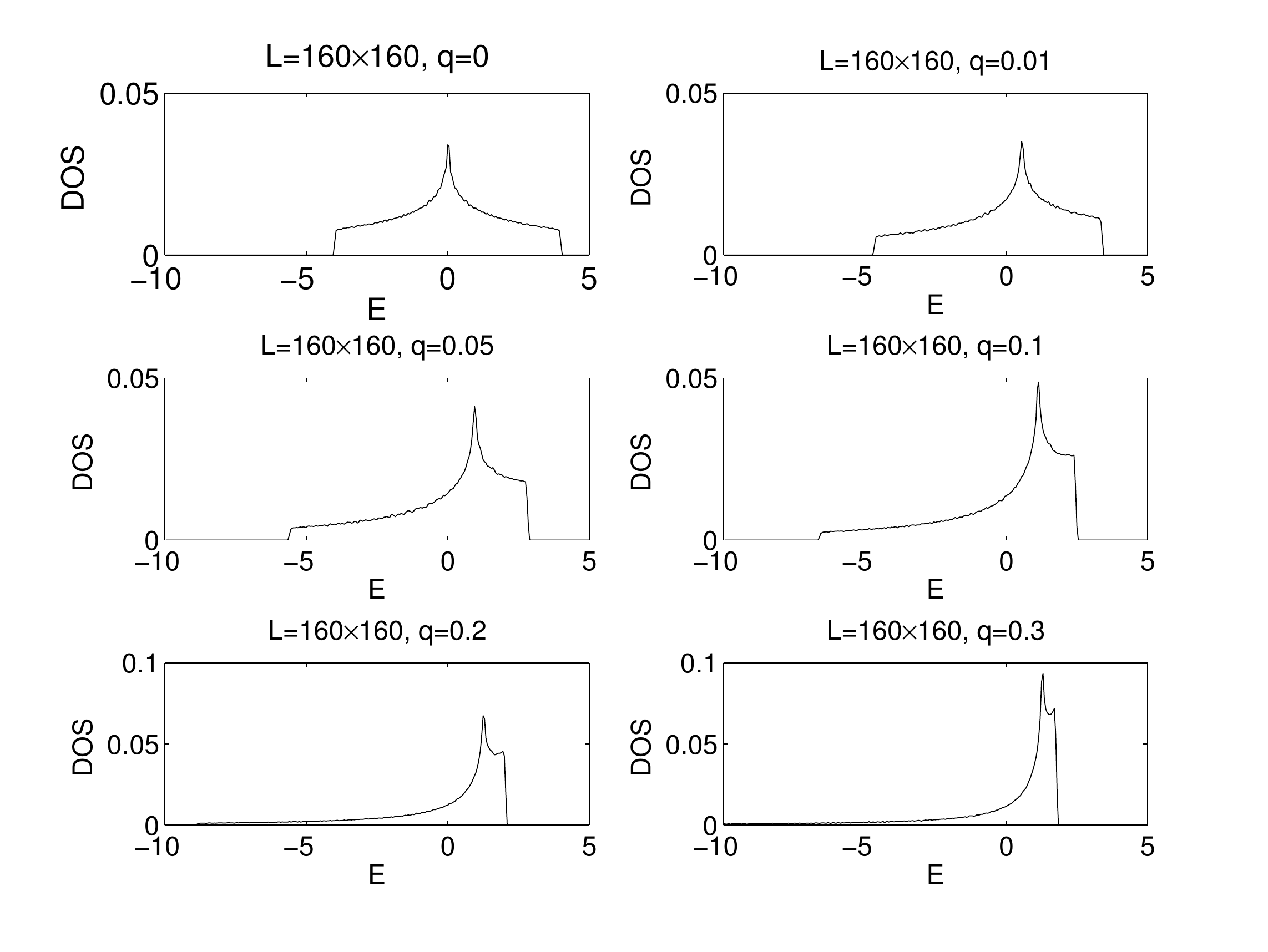}
\end{center}
\caption{The non-interacting $(D=2)$ DOS corresponding to the long-range
hopping for different $q$ and $L=160\times160$~\cite{Fark2}.}
\label{fig3}
\end{figure}

\begin{figure}[!t]
\begin{center}
\includegraphics[width=7cm]{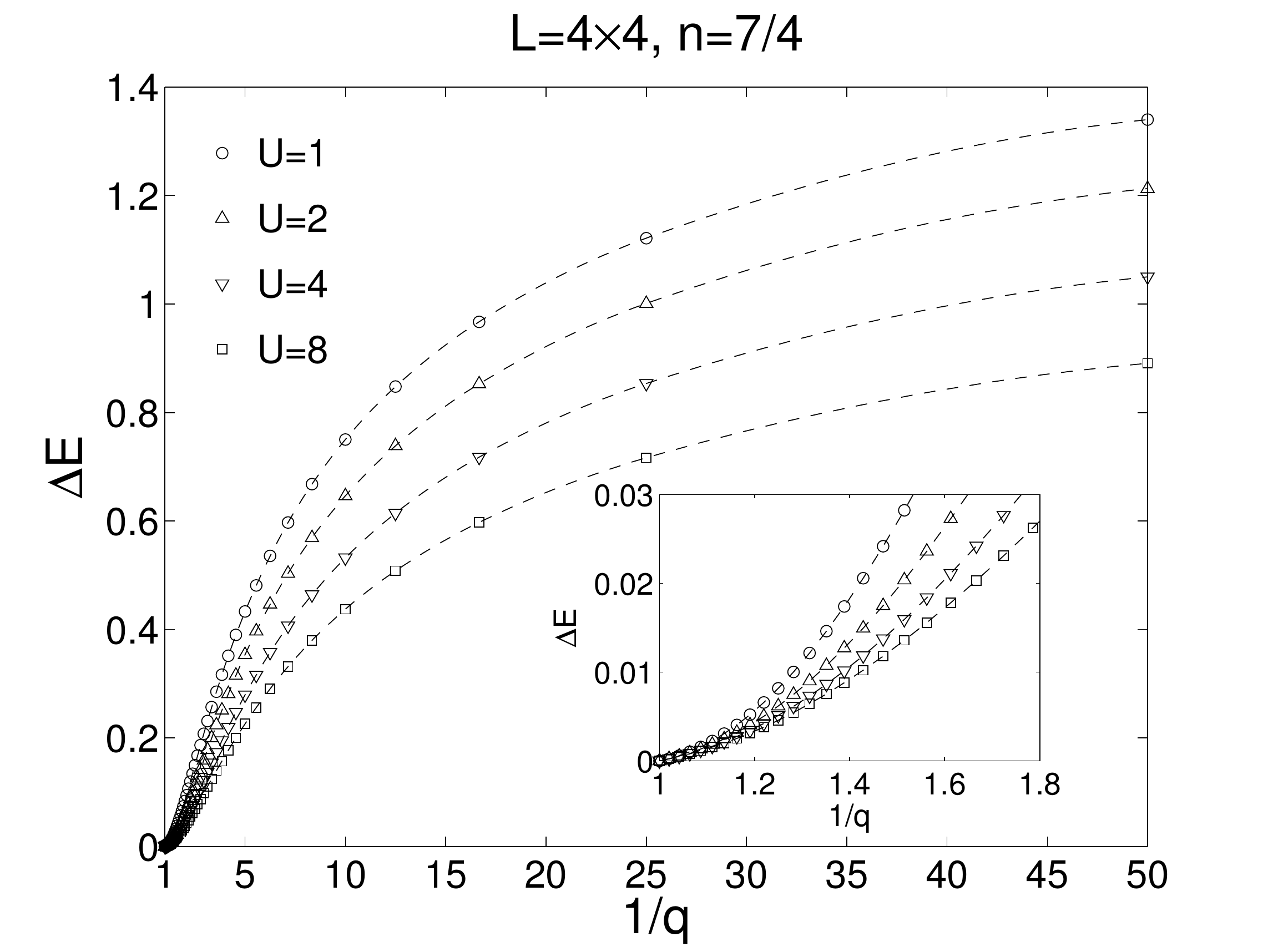}
\includegraphics[width=7cm]{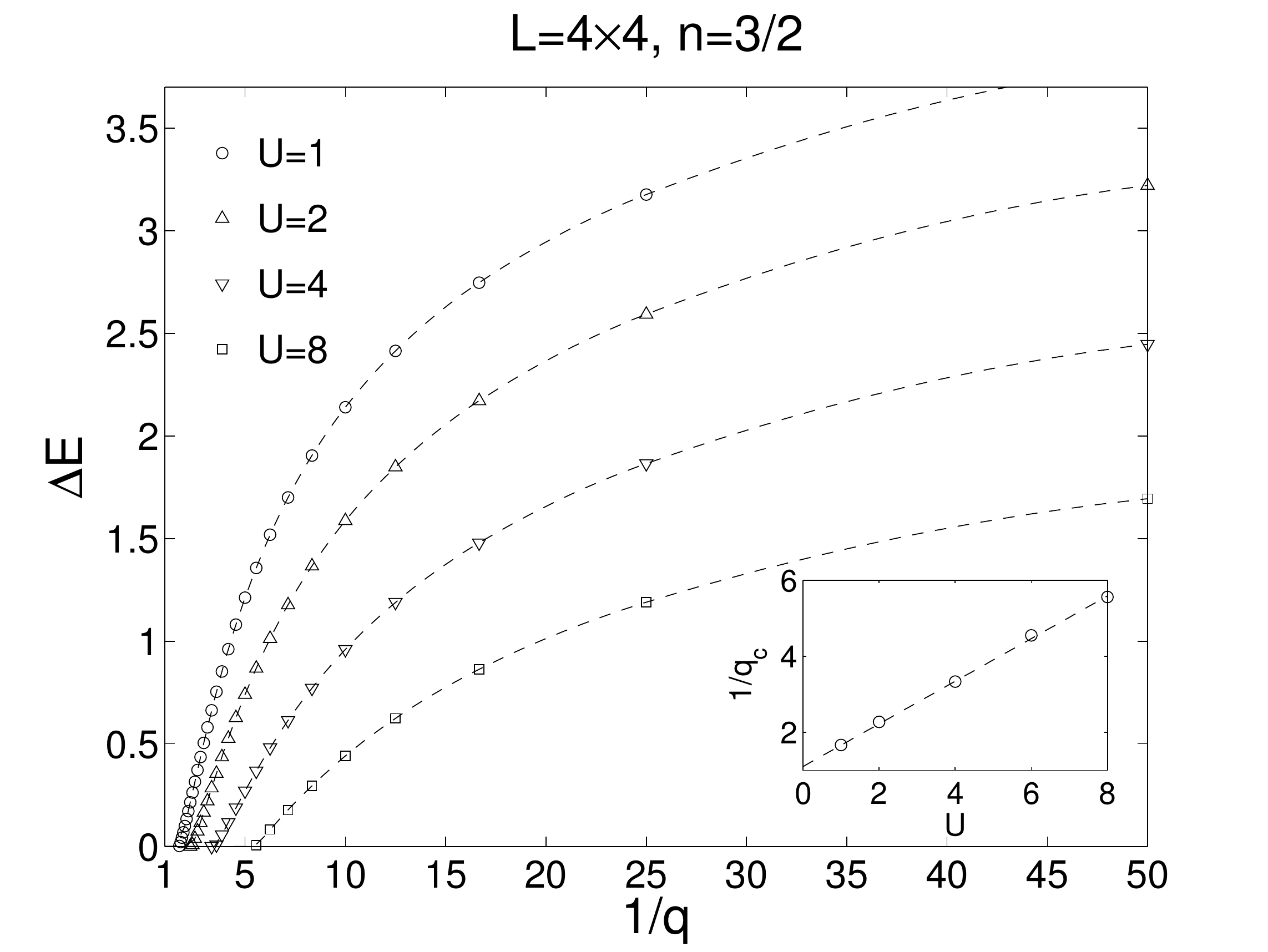}
\end{center}
\caption{ The left-hand panel: The difference 
$\Delta E=E_f-E_g$ between the exact ground state $E_g$ and
the ferromagnetic state $E_f$ (which is exactly known
since the state has no double hole occupancy)
as a function of $1/q$ calculated for 
$n=7/4$ and different $U$ on the $L=4\times4$ cluster.  The inset
shows the situation in the limit $1/q \to 1$. The right-hand panel: 
$\Delta E=E_f-E_g$  as a function of $1/q$ calculated for 
$n=3/2$ and different $U$ on the $L=4\times4$ cluster. The inset shows
the $U$ dependence of $1/q_c$~\cite{Fark2}.}
\label{fig4}
\end{figure}

Similarly to the one-dimensional case, let us start with a discussion
of long-range hopping effects on the noninteracting two-dimensional DOS
(see figure~\ref{fig3}). It is seen that with an increasing 
$q$, more weight shifts to the upper edge of the band and the DOS becomes 
strongly asymmetric, indicating possible ferromagnetic regions
in the limit $n>1$ 
For this reason, we have focused  our attention on the case
of electron concentrations above half-filling $n>1$.
First, we have examined the model Hamiltonian by the exact diagonalization 
method on the finite $4 \times 4$ cluster for several selected values
of on-site Coulomb interaction $U$ ($U=1,2,4,8$) and electron 
concentrations $n > 1$ ($n=3/2$ and $n=7/4$).  The results of our numerical 
calculations are displayed in figure~\ref{fig4}. 
There is plotted the difference
$\Delta E=E_f-E_g$ between the exact ground state $E_g$ and
the ferromagnetic state $E_f$ (which is exactly known) as a function of $1/q$.
The ground state is ferromagnetic in the regions
where $\Delta E=0$. It can be seen that for higher electron concentrations ($n=7/4$),
the ground states of the two-dimensional Hubbard model with generalized
hopping are non-ferromagnetic for all examined values of the on-site 
interaction $U$, which strongly contrasts with the one-dimensional
case, where the ferromagnetic state has been stabilized for all
electron concentrations above the half-filled band case $n>1$.
However, for smaller values of electron concentrations ($n=3/2$),
the situation is fully different. In this case, the ferromagnetic
state is the ground state of the model for all examined values of
$U$ above some critical value of the long-range hopping parameter
$q_c$. As shown in the inset in
figure~\ref{fig4}, $1/q_c$ scales linearly with $U$, from which it can be
directly determined 
that $1/q_c=1.13+0.558U$. Analysing these results one can 
see that already for relatively small values of the Coulomb interaction
$U$, the critical values of the long-range hopping parameter $q_c$ are
from the physically realistic regime (e.g., $q_c\cong 0.44$ for $U=2$,
$q_c\cong 0.3$ for $U=4$ and $q_c\cong 0.18$ for $U=8$). This confirms the
importance of the long-range electron hopping term for a correct description
of ferromagnetism in real materials.

\begin{figure}[!t]
\begin{center}
\includegraphics[width=7cm]{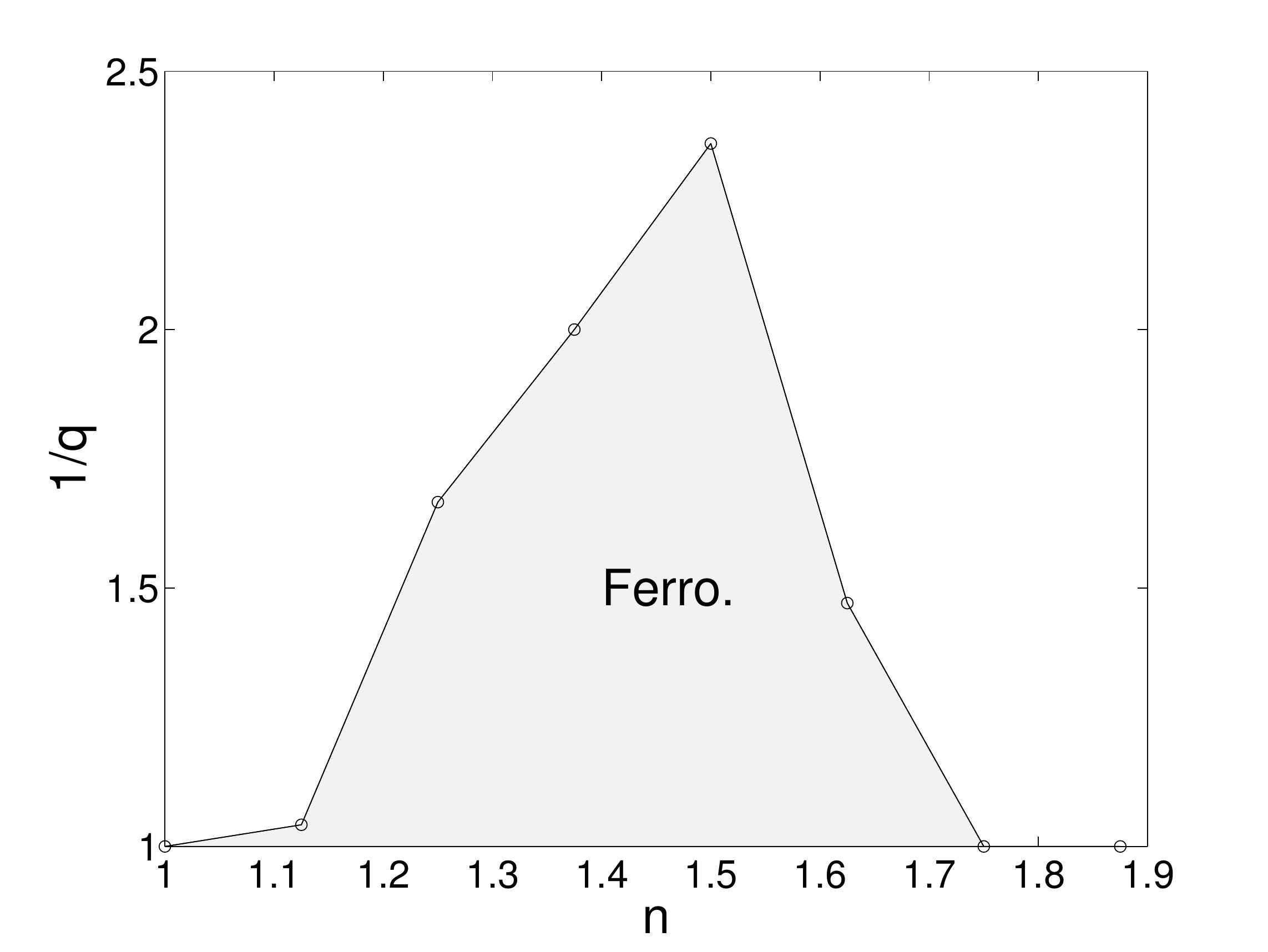}
\end{center}
\caption{The phase diagram of the two-dimensional Hubbard model
with long-range hopping in the $1/q-n$ plane calculated for $U=2$
and $L=4\times4$~\cite{Fark2}.}
\label{fig5}
\end{figure}

Since our numerical results revealed a strikingly different behaviour
of the model in one and two dimensions for electron concentrations
above half filling, and namely, the existence of the critical electron
concentration $n_c$ below (above) which the ground state is ferromagnetic
(non-ferromagnetic), we have decided to study in detail the comprehensive
phase diagram of the model in the $q$-$n$ plane. Again we have used the small
cluster exact diagonalization technique and the cluster of $L=4\times4$
sites. The results of our numerical calculations obtained for the
intermediate value of the Coulomb interaction $U=2$ are summarized in
figure~\ref{fig5}.
They show that just $n=7/4$ is the critical electron concentration 
above which the ground state is non-ferromagnetic. Below this value, 
the ground state is ferromagnetic. The critical value of the inverse 
long-range hopping parameter $1/q_c$ increases with an decreasing $n$ and
reaches its maximum ($1/q_c \sim 2.4$) at $n=3/2$. A further decrease
in $n$ gradually reduces the critical value of $1/q_c$ to one.
A similar behaviour of the model has been observed for both smaller
($U=1$) as well as larger ($U=4$) values of the Coulomb interaction.
Thus, we can conclude that the ferromagnetic phase in the two dimensional
Hubbard model, in spite of its partial reduction (in comparison to the
one dimensional case), remains robust.

Unfortunately, these results cannot be considered as definite, since
they were obtained on a very small cluster and, therefore, it is
necessary to prove them independently by other methods. To fulfill
this goal, we  performed the same calculations for one representative
value of $U$ ($U=2$) and for two representative values of $n$ ($n=3/2\,,5/4$)
by the projector quantum Monte Carlo (QMC) method~\cite{Sorella,Loh,Imada}
that is capable of treating several times larger clusters with high accuracy.
The QMC simulations were performed using a projector algorithm
which applies $\exp(-\theta H)$ to a trial wave-function (in our case,
the solution for $U=0$). A projector parameter $\theta \sim 30$ and a
time slice of $\Delta \theta = 0.05$ suffice to reach well converged values 
of the observables discussed here. 

\begin{figure*}[!t]
\begin{center}
\includegraphics[width=5.5cm]{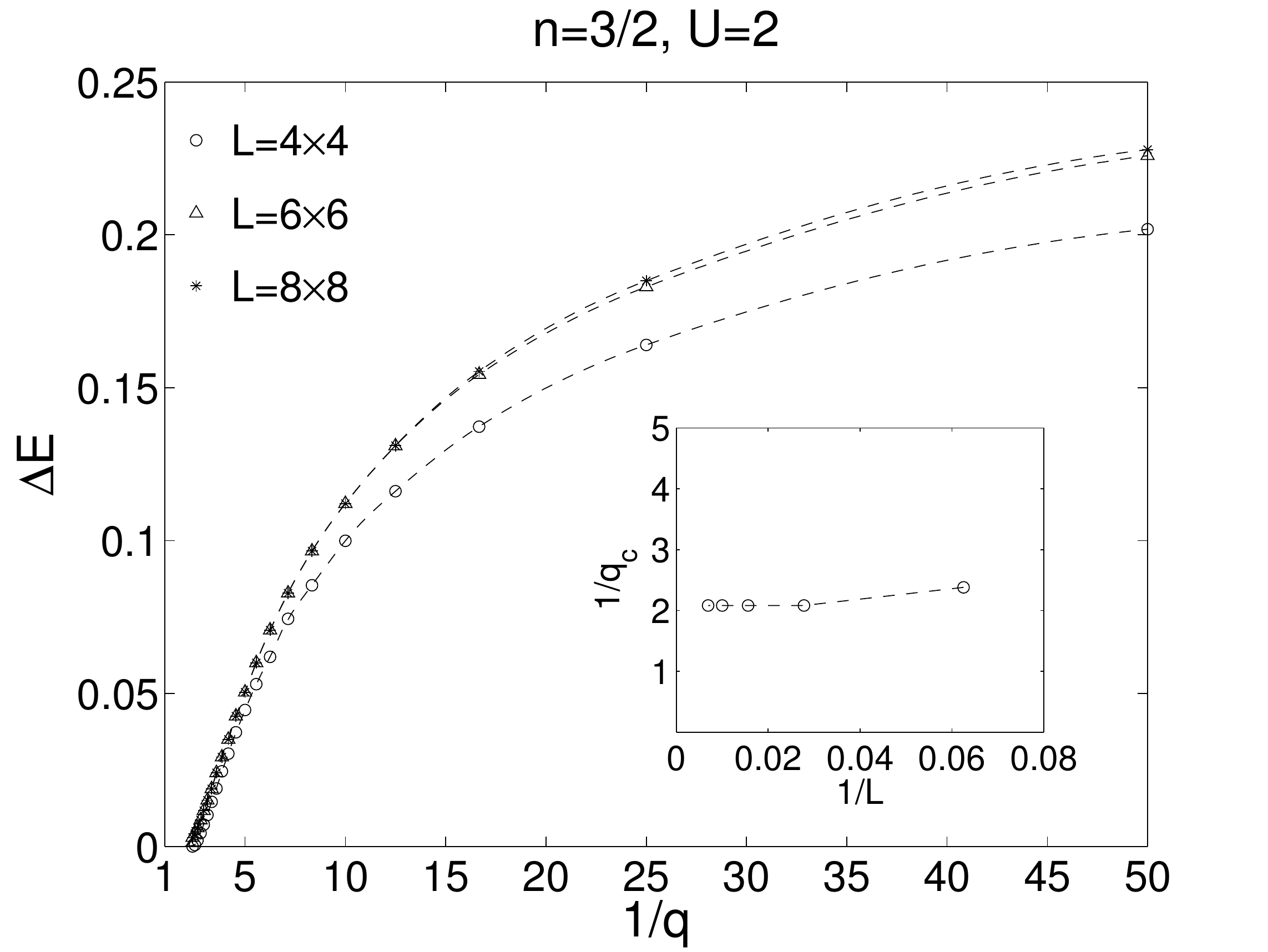}
\includegraphics[width=5.5cm]{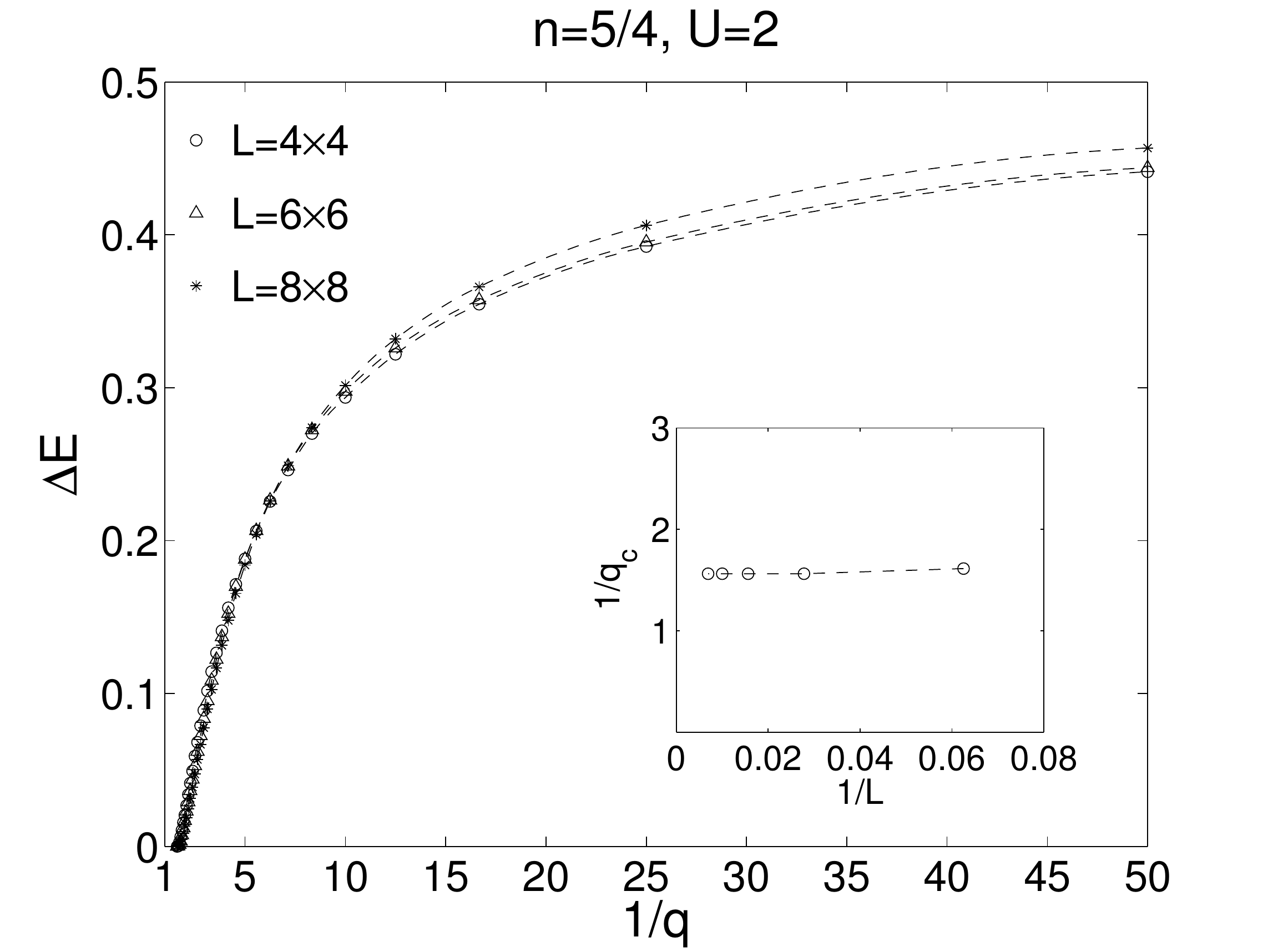}
\end{center}
\caption{ The difference $\Delta E=E_f-E_g$ between the ground 
state $E_g$ and the ferromagnetic state $E_f$ as a function of $1/q$ 
calculated for $U=2$, two different values of $n$ ($n=3/2$ and $n=5/4$)
and three different finite clusters of $L=4\times4$, $L=6\times6$ and
$L=8\times8$. The QMC results~\cite{Fark2}.
}
\label{fig6}
\end{figure*}

The results of numerical calculations performed  on larger clusters
consisting of $L=6\times 6$ and $L=8\times 8$ sites (over the whole interval $[0,1]$
of $q$ values with the step $\Delta q=0.02$) and clusters of
$L=10\times 10$ and $L=12\times 12$ sites (over a restricted set of $q$
values near $q_c$) are displayed in figure~\ref{fig6}. 
\vspace{-1mm}  
These results clearly show that the
finite size effects on the critical values of the long-range hopping parameter
are negligible (see insets in figure~\ref{fig6}), and thus, the magnetic phase diagram found for the $4\times 4$
cluster can be satisfactorily extrapolated on macroscopic systems. 

\begin{figure}[!t]
\begin{center}
\includegraphics[width=6.6cm]{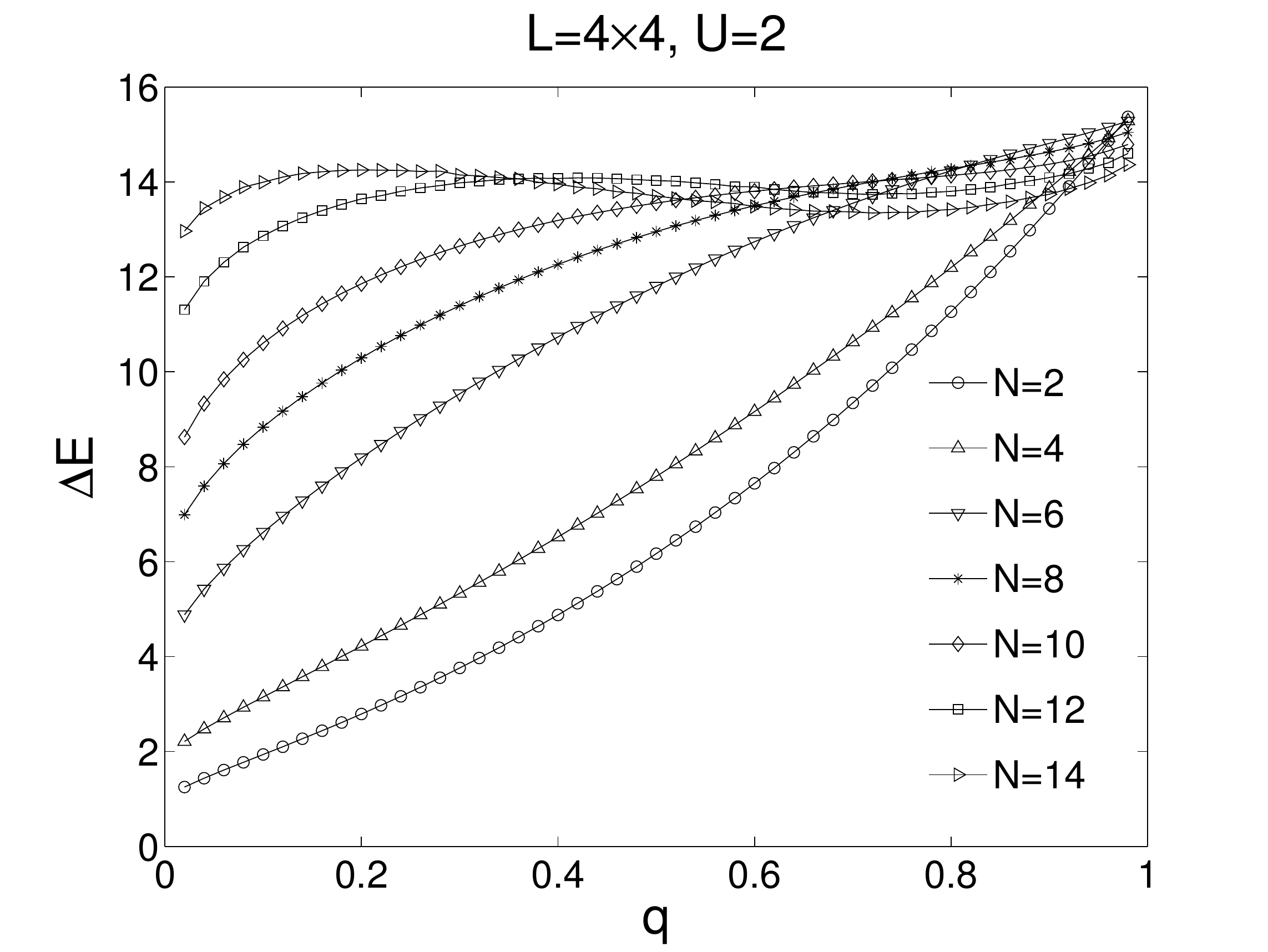}
\end{center}
\caption{ The difference
$\Delta E=E_f-E_g$ between the exact ground state $E_g$ and
the ferromagnetic state $E_f$ as a function of $q$ calculated for
$U=2$ and different electron fillings $N$ on the $L=4\times4$ cluster~\cite{Fark2}.}
\label{fig7}
\end{figure}

Finally, let us briefly discuss the numerical results obtained for electron
concentrations less than the half-filled band case ($n=1$). They are displayed
in figure~\ref{fig7} for $U=2$ and the complete set of even electron fillings
with $N<L$ on the $4\times 4$ cluster.
These results clearly demonstrate the absence of
the ferromagnetic ground state in the two-dimensional Hubbard model with
exponentially decaying hopping amplitudes for all values the hopping parameter $q$
and electron concentrations $n<1$, showing a key role of the band filling
$n$ in the mechanism of stabilization of the ferromagnetic state.  

\subsection{The effect of correlated hopping}

Let us now further generalize the single-band Hubbard model with long-range
hopping by introducing the correlated hopping 
term~\cite{Shvaika1,Shvaika2,farky_ch1,farky_ch2}, 
in which the $\sigma$-electron 
hopping amplitudes between lattice sites $i$ and $j$ depend explicitly on $n_{i-\sigma}$ 
and $n_{j-\sigma}$ occupancy, i.e., 
\begin{equation}
t^{\sigma}_{ij}=t_{ij}[1+t'(n_{i-\sigma}+n_{j-\sigma})].
\label{Eq4}
\end{equation}
The importance of the action of the correlated hopping term on the ground-state 
properties of the Hubbard model has been already mentioned by 
Hubbard~\cite{Hubbard}. Later, Hirsch~\cite{Hirsch} pointed
out that this term may be relevant in explaning the
superconducting properties of strongly correlated electrons.
Here, we discuss the effects of this term on the stability of 
the fully polarized ferromagnetic state. The same subject was 
studied by Amadon and Hirch~\cite{Amadon}, as well as by
Kollar and Vollhardt~\cite{Kollar}, though 
they considered the hopping only between the nearest-neighbor sites.

The Hamiltonian of the single-band Hubbard model in which the effects 
of long-range and correlated hopping are incorporated is given by

\begin{equation}
H=\sum_{ij\sigma}t^{\sigma}_{ij}c^+_{i\sigma}c_{j\sigma}+
U\sum_{i}n_{i\uparrow}n_{i\downarrow}.
\label{Eq5}
\end{equation}

\begin{figure*}[!t]
\begin{center}
\includegraphics[width=6.0cm]{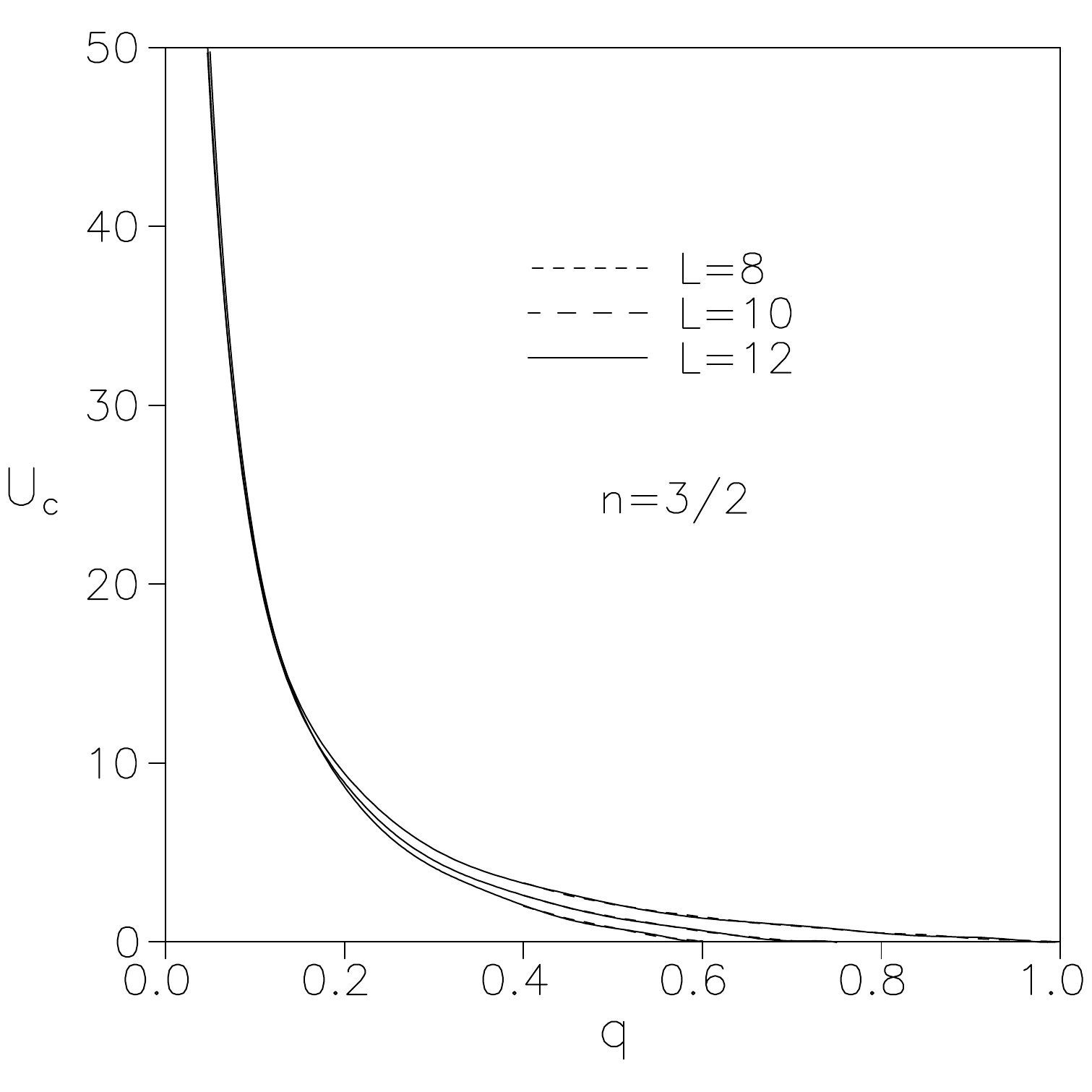}
\includegraphics[width=6.0cm]{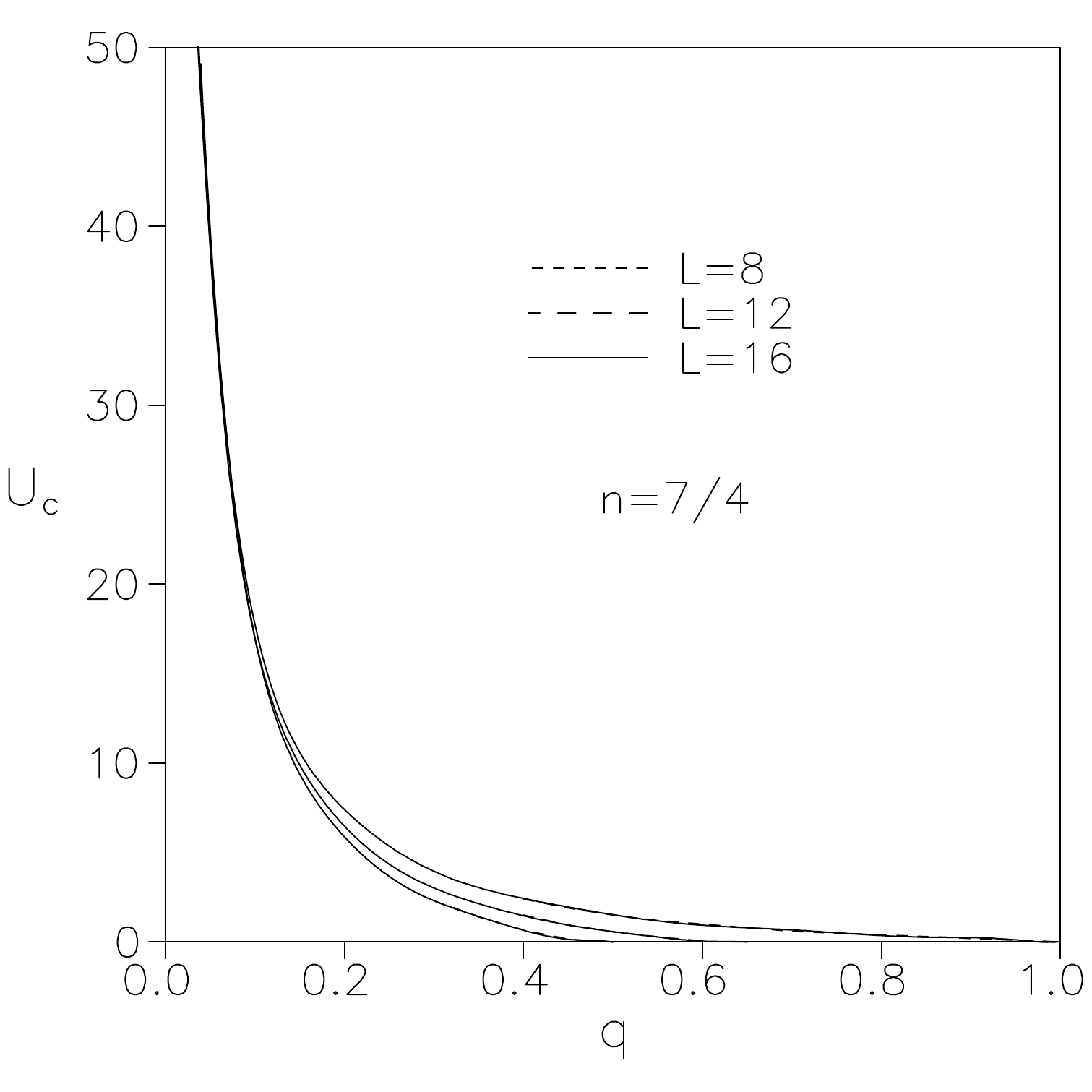}
\end{center}
\caption{The critical interaction strength $U_{c}$  as a function 
of $q$ calculated for different $t'$ and $L$ at $n=3/2$ (the left-hand panel)
and at $n=7/4$ (the right-hand panel). 
Curves from up to down correspond to: $t'=0, 0.2$ and 0.4~\cite{Fark4}.}
\label{fig8}
\end{figure*}

To examine the possibilities for the existence of ferromagnetism in this model,
the ground states are determined by exact diagonalizations for a 
wide range of model parameters ($q,t',U,N=\sum_{\sigma}N_{\sigma} $). 
Typical examples are then chosen from a large number of available 
results to represent the most interesting cases.  
The results of our small-cluster exact-diagonalization calculations
obtained on finite clusters up to $L=16$ sites are summarized in 
figure~\ref{fig8}. 
There is shown the critical interaction strength $U_c$, 
above which the ground state is ferromagnetic, as a function of $q$ for the
selected values of $n=N/L$ and $t'$ ($n=3/2\,,7/4; t'=0,0.2\,,0.4$). 
To reveal the finite-size effects on the stability of ferromagnetic domains, 
the behavior of the critical interaction strength $U_c(q)$ was calculated
on several finite clusters at each electron filling. It is seen that 
finite-size effects on $U_c$ are small and thus these results can be 
satisfactorily extrapolated to the thermodynamic limit $L\to \infty$.           
Our results clearly demonstrate that the ferromagnetic state is strongly 
influenced by correlated hopping ($t'$) and generally it is stabilized with 
increasing $t'$. The effect is especially strong for intermediate and strong 
values of $q$. There even exists some critical value of $q$ above which 
the ground state 
is ferromagnetic for all nonzero $U$. With an increasing $t'$, this critical
value shifts to lower values of $q$ (that represent a much more realistic 
type of electron hopping) and the ferromagnetic domain correspondingly
increases. Performing exhaustive numerical studies of the model for a wide
range of electron concentrations (on different lattice clusters) we have
found that the model exhibits the same behavior for all electron
concentrations above half-filling\footnote{$n\leqslant 1$ and $t'<0$ does not stabilize the ferromagnetic 
state.}
and that with an increasing
concentration this effect becomes more pronounced.
These results clearly show that ferromagnetism comes naturally from the 
Hubbard model with long-range and correlated hopping for a wide range of model 
parameters without any other assumptions.

\subsection{The effect of long-range Coulomb interaction}

The microscopic model of electronic correlations in solids based on the picture 
of long-range electron hopping with exponentially decaying hopping 
amplitudes in  combination with the standard on-site description of the Coulomb 
interaction between electrons on a lattice does not model very realistically the 
situation in real transition metal compounds. Indeed, it would be more correct to
describe the Coulomb interaction between electrons by a similar formula
that corresponds to the electron hopping [equation~(\ref{Eq2})],
i.e., by exponentially decaying interaction amplitudes $U_{ij}$,
\begin{equation}
U_{ij}(q_U)=\left \{ \begin{array}{ll}
  U,               &   i=j,\\\\
\displaystyle{\frac{U}{2}q_U^{|i-j|}},          &   i\neq j,
\end{array}
\right.
\label{Eq6}
\end{equation}
where $U$ is the strength of the on-site Coulomb interaction and $q_U$
($ 0 \leqslant q_U \leqslant 1$) is the parameter of the long-range Coulomb interaction.
Such a selection is natural, since the overlap of the atomic wave functions
decreases exponentially with a distance of two atoms on a lattice and thus
the model based on such a supposition describes more realistically the 
situation in transition metal compounds. 

\begin{figure}[h!]
\begin{center}
\includegraphics[width=7.6cm]{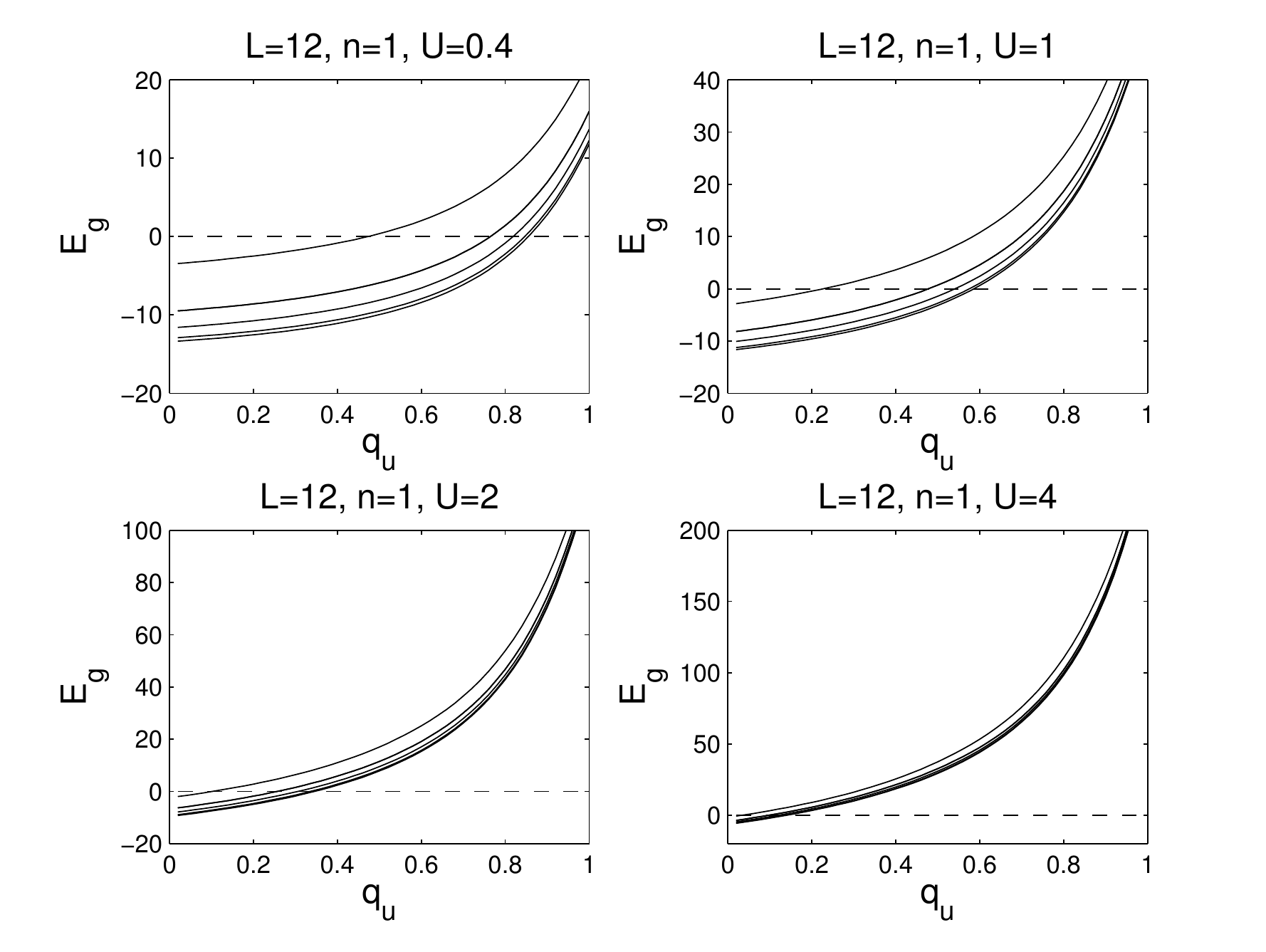}
\includegraphics[width=7.6cm]{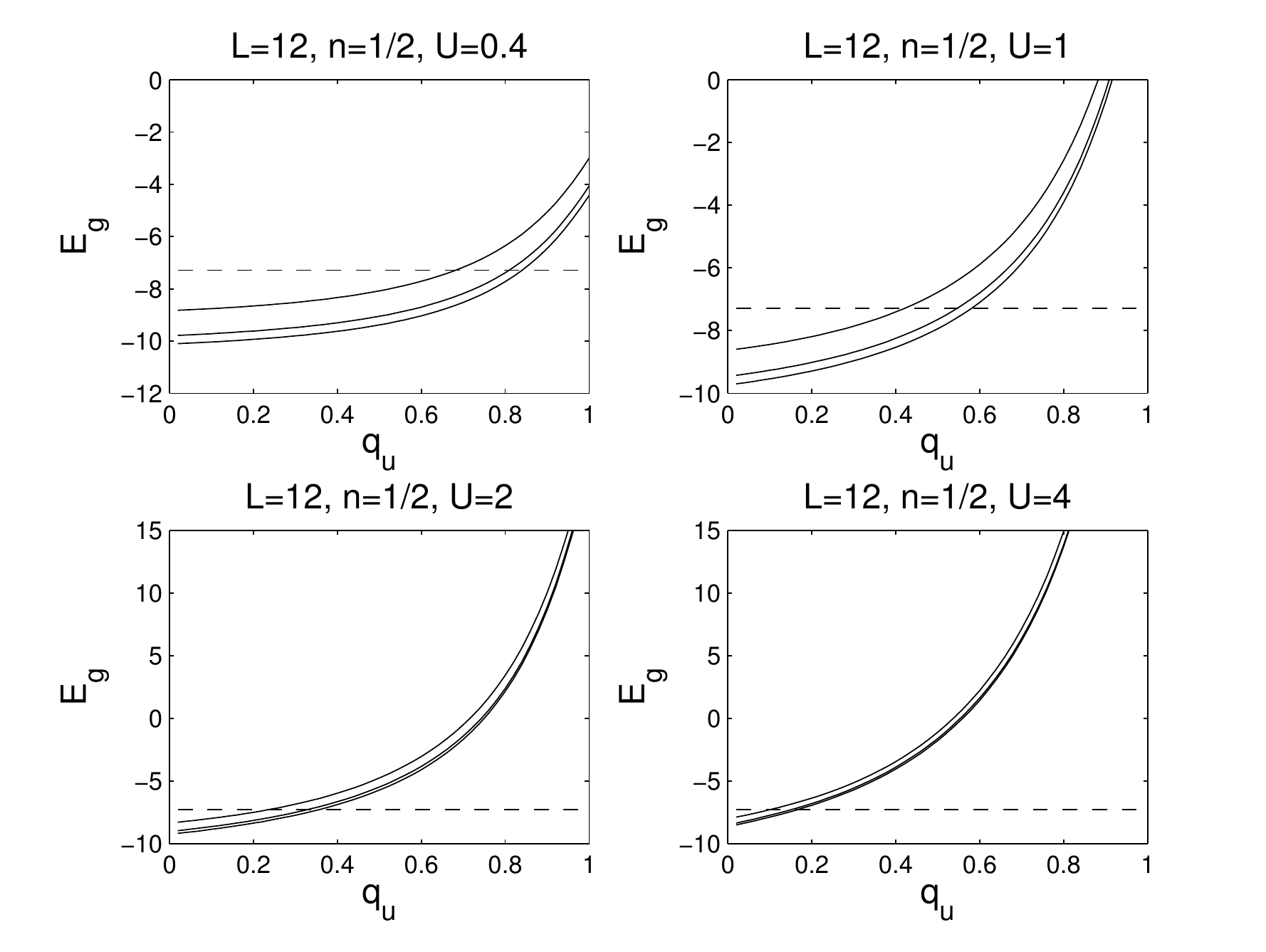}
\end{center}
\caption{The left-hand panel: The ground-state energy $E_g$ as a function of $q_U$ 
calculated for $q_t=0, L=12, n=1$ and different values of $U$ and $N_{\uparrow}$. 
The dashed line corresponds to $N_{\uparrow}=0$ (the fully polarized state)
and the full lines (from up to down) correspond to $N_{\uparrow}=1,2,3,4,5$
(the partially polarized states) and $N_{\uparrow}=6$ (the non-polarized
state). The right-hand panel: The ground-state energy $E_g$ as a function of $q_U$ calculated for 
$q_t=0, L=12, n=1/2$ and different values of $U$ and $N_{\uparrow}$. 
The dashed line corresponds to $N_{\uparrow}=0$ (the fully polarized state)
and the full lines (from up to down) correspond to $N_{\uparrow}=1,2$
(the partially polarized states) and $N_{\uparrow}=3$ (the non-polarized
state)~\cite{Fark3}.    
}
\label{fig9}
\end{figure}

Thus, our generalized Hamiltonian taking into account the effects of the
long-range electron hopping as well as the effects of the long-range
Coulomb interaction, both in the form of exponentially 
decaying amplitudes, can be written as follows:
\begin{equation}
H=\sum_{ij\sigma}t_{ij}(q_t)c^+_{i\sigma}c_{j\sigma}+
 \sum_{ij\sigma\sigma'}U_{ij}(q_U)n_{i\sigma}n_{j\sigma'}.
\label{Eq7}
\end{equation}
To reveal the behaviour of the model in different concentration limits,
we  selected three different values of the electron concentration, 
and namely, $n=1/2$, $n=1$ and $n=3/2$ that represent three physically 
most interesting limits of the model. Let us start the discussion of our 
results with the half-filled band case $n=1$. To see the pure effects of 
the long-range Coulomb interaction on the stability of the ferromagnetic
state, we  first examined the conventional limit of the electron hopping
corresponding to the following selection of the hopping amplitudes $t_{ij}=-1$ 
if $i$ and $j$ are the nearest neighbour and $t_{ij}=0$ otherwise. For this case,
we  calculated the ground states of the model $E_g(N_{\uparrow})$ in all
different spin sectors $S_z=N_{\uparrow}-N_{\downarrow}$ as a function of 
$U$ and $q_U$ with steps $\Delta U=0.2$ and $\Delta q_U=0.02$. The resulting
behaviours of $E_g(N_{\uparrow})$  as functions of $q_U$ and $U$ are
used directly to identify the stability regions of the fully polarized state,
i.e., where the following inequality $E_g(0) < \textrm{min}[E_g(1),E_g(2), \dots,
E_g(N/2)]$ is satisfied for a given $U$ and $q_U$. 
The results of our numerical simulations obtained for $L=12, n=1$ and four
different values of $U$ are shown in figure~\ref{fig9} (the left-hand panels). 
One can see that for each selected value of the Coulomb interaction $U$, there
exists a critical value of the interaction parameter $q_U$ above which the
ground state is the fully polarized state, i.e., $N_{\uparrow}=0$, 
$N_{\downarrow}=L$. Since the ground state of the ordinary (nearest-neighbour)
half-filled Hubbard model in  one dimension is antiferromagnetic for
all Coulomb interactions, our results point to the crucial role of the 
long-range interaction on the stabilization of the ferromagnetic
state. Moreover, we have found that the critical value of $q_U$ above which
the ground state is ferromagnetic decreases rapidly with an increasing $U$
and thus already small, but physically the most realistic values of $q_U$,
are capable of generating the ferromagnetic state for intermediate and strong 
Coulomb interactions. The same conclusions are also valid  for electron 
concentrations smaller than the half-filled band case ($n<1$), as illustrates
figure~\ref{fig9} (the right-hand panels) for the quarter-band filling $n=1/2$, while 
in the opposite limit $n>1$ no effects of long-range interaction on the stabilization 
of the band ferromagnetism are observed.
Comparing these results with the ones discussed above 
for $q_U=0$ and $q_t > 0$, one can find a fundamentally different response
of the electron system to the long-range electron hopping and the 
long-range Coulomb interaction. Indeed, while the long-range electron
hopping itself, i.e., without the long-range Coulomb interaction, stabilizes
the ferromagnetic state for the electron concentrations above the half-filled
band case ($n>1$), the long-range Coulomb interaction itself stabilizes
the ferromagnetic state in the opposite limit, i.e., for $n \leqslant 1$. 
Thus, one can intuitively expect that the combined effects of the long-range
electron hopping and the long-range Coulomb interaction could lead to the 
stabilization of the ferromagnetic state for the electron concentrations smaller
as well as larger than the half-filled band case. 

\begin{figure}[!t]
\begin{center}
\includegraphics[width=5.2cm]{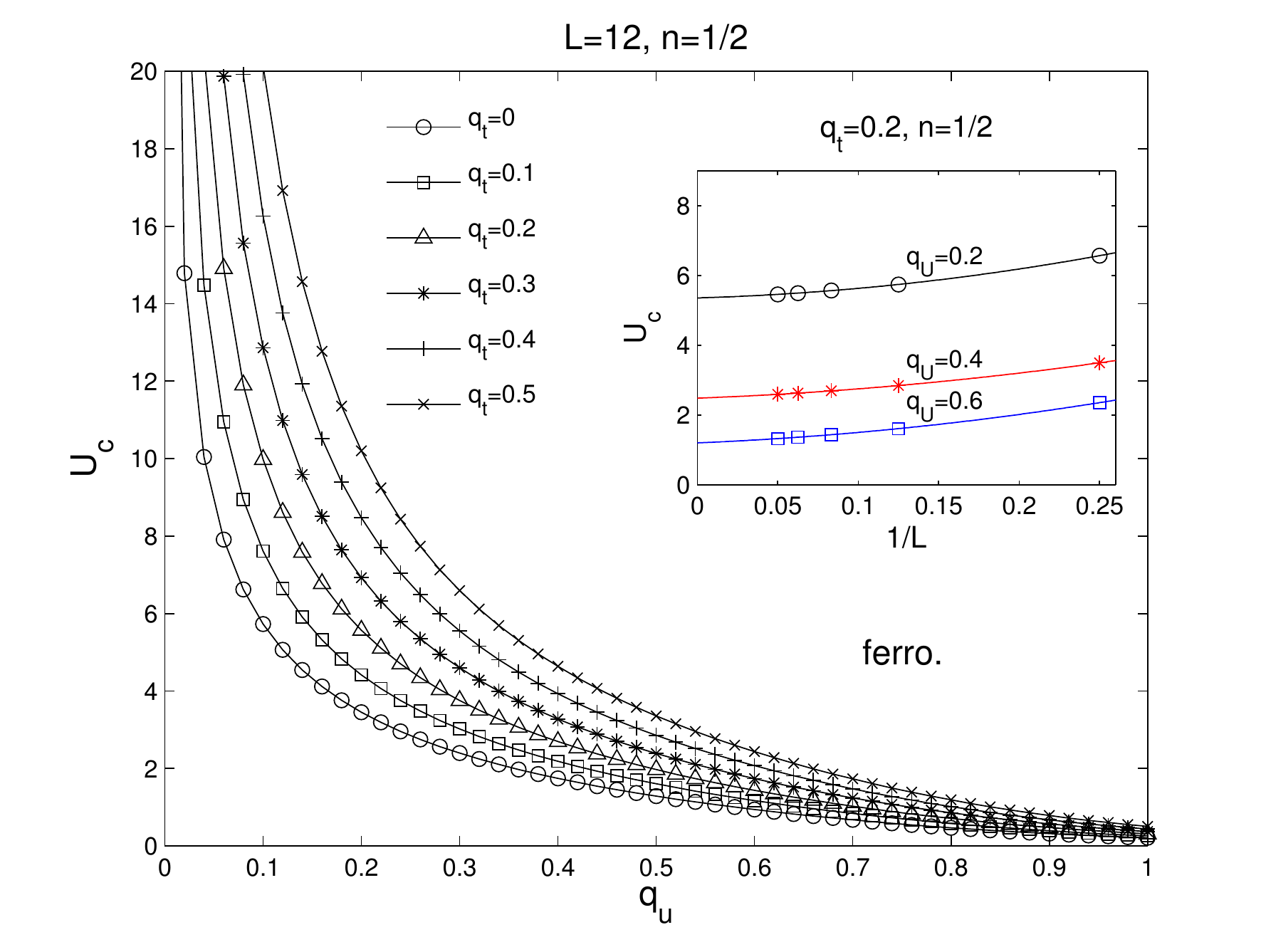}
\includegraphics[width=5.2cm]{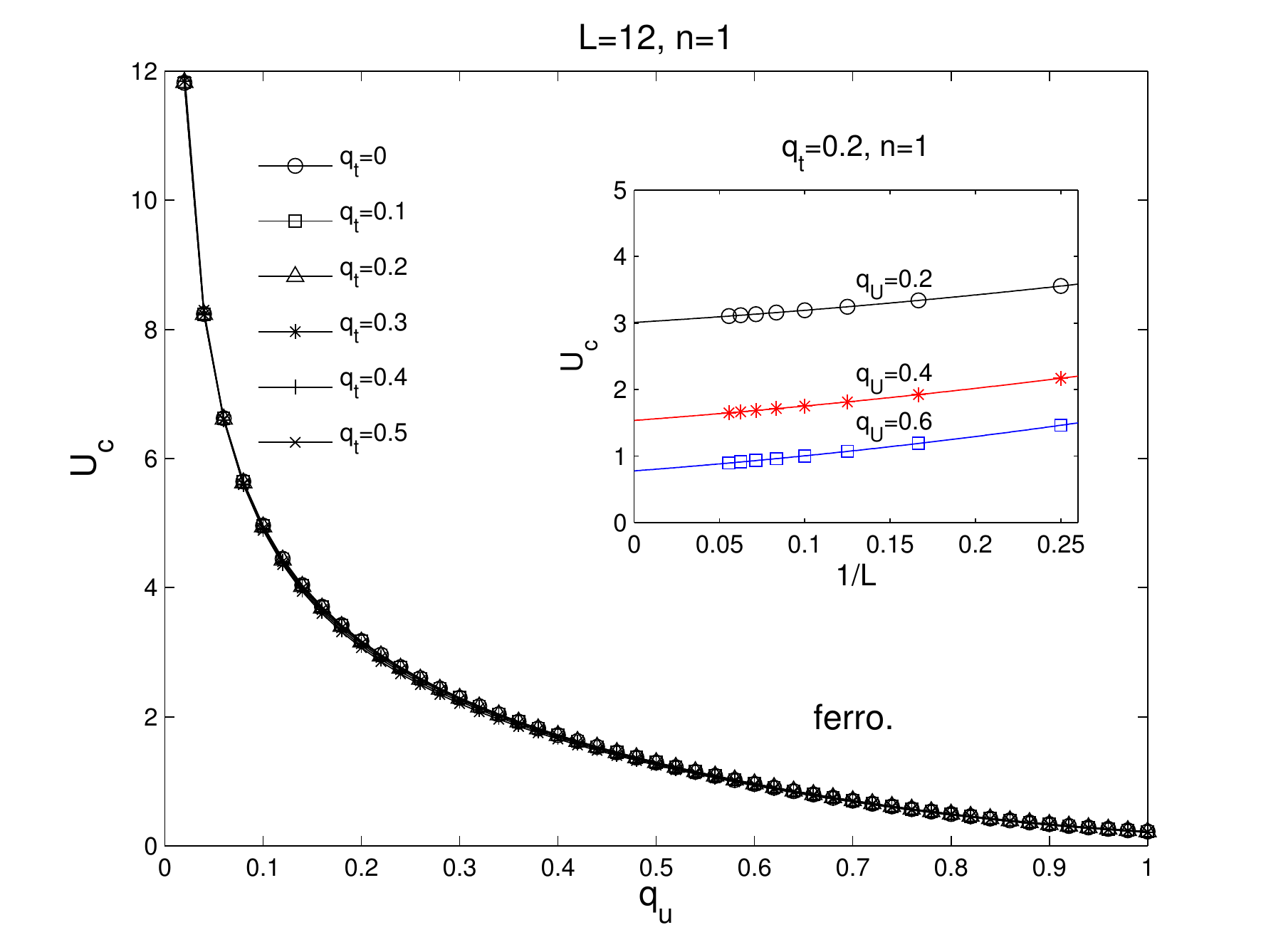}
\includegraphics[width=5.2cm]{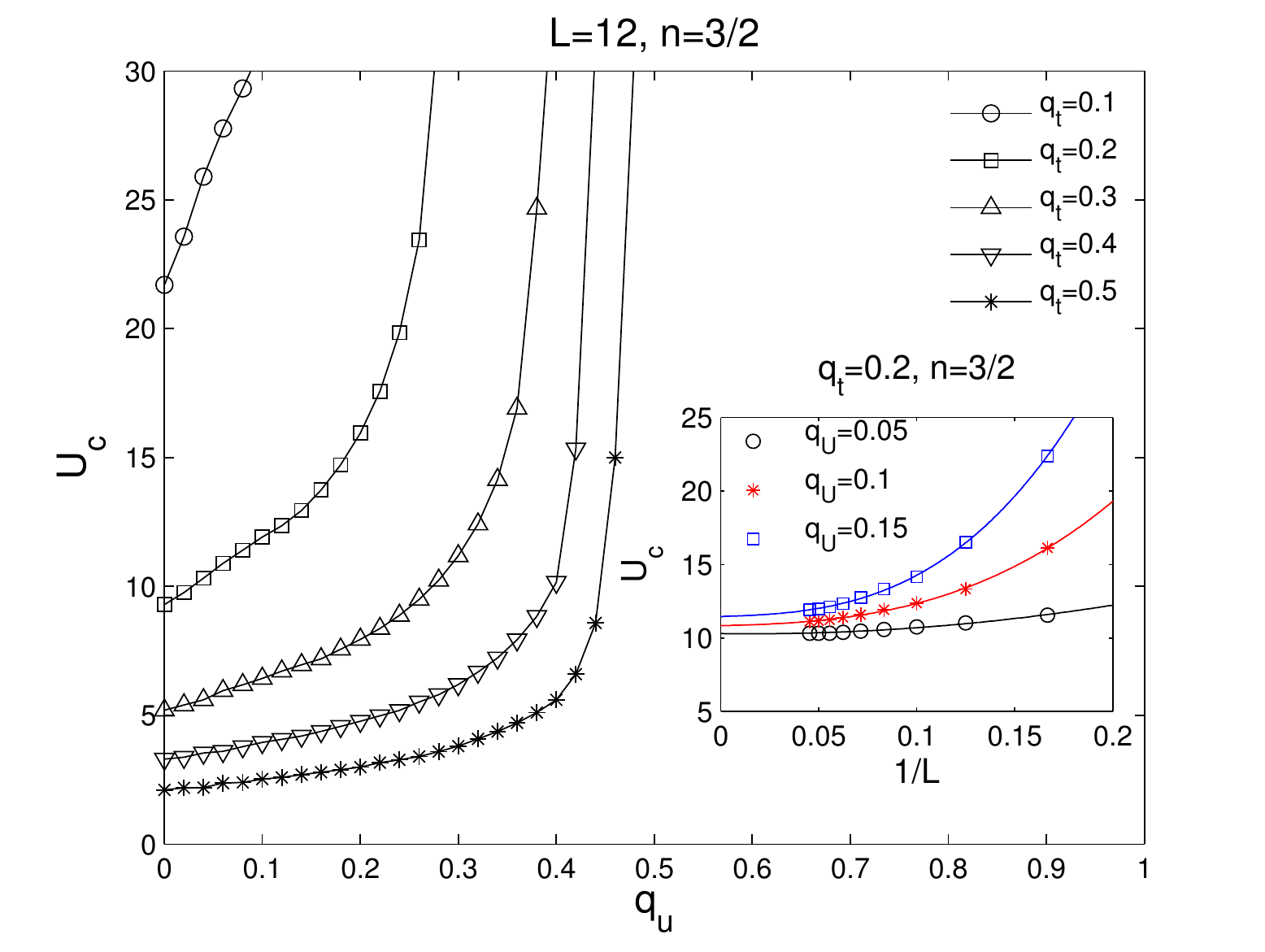}
\end{center}
\caption{(Colour online) The left-hand panel: The critical interaction strength $U_c$ as a function
of $q_U$ calculated for $L=12, n=1/2$ and different values of $q_t$. The inset shows the
scaling of $U_c$ calculated for three different values of $q_U$. The scaling function
is the second order polynomial expression in $1/L$.
The middle panel: $U_c$ as a function
of $q_U$ calculated for $L=12, n=1$ and different values of $q_t$.
The inset shows scaling $U_c$ calculated for three different values of $q_U$. 
The scaling function is the second order polynomial expression in $1/L$.
The right-hand panel: $U_c$ as a function of $q_U$ calculated for $L=12, n=3/2$ 
and different values of $q_t$. The inset shows the scaling of $U_c$ 
calculated for three different values of $q_U$. The scaling function
is the third order polynomial expression in $1/L$~\cite{Fark3}.
}
\label{fig10}
\end{figure}

To verify this conclusion, we  performed exhaustive numerical
studies of the model for a wide range of the model parameters. In particular,
we  selected five different values of the long-range hopping parameter 
$q_t >0$ ($q_t=0.1, 0.2, 0.3, 0.4, 0.5$) and for each of them we  calculated
the ground-state energy $E_g(N_{\uparrow})$ in all different spin sectors,
for the following sequences of $q_U$ and $U$ values: $q_U=0,0.02, \dots, 1$ 
and $U=0,0.2, \dots, 30$. The results obtained are summarized in the form of
the $U$--$q_U$ magnetic phase diagram, where the critical interaction 
strength $U_c(q_U)$ represents the phase boundary above which the ground
state is a fully polarized state. The situation for the quarter-band
filling is shown in figure~\ref{fig10} (the left-hand panel).
Our results clearly demonstrate that the
ferromagnetic state is strongly influenced by $q_U$ for electron 
concentrations bellow half-filling and generally it is stabilized with an
increasing $q_U$. The effect is especially strong for small values of $q_U$
where small changes of $q_U$ reduce dramatically the critical interaction
strength $U_c$ and so the ferromagnetic state becomes stable for a wide range
of Coulomb interaction $U$. The second general trend that can be deduced
from our numerical calculations is depicted by the ferromagnetic domain being reduced
with an increasing $q_t$, though it remains robust for a wide range of $q_U$ and
$U$ values. For $n<1$, this result is intuitively expected since the
long-range hopping itself stabilizes the ferromagnetic state only for 
$n>1$. To exclude the finite-size effects on the stability 
of ferromagnetic domains we  performed an exhaustive numerical study of
the $L$-dependence of the critical interaction strength $U_c$ for selected
values of $q_t$ and $q_U$ on finite clusters from $L=4$ to $L=20$.  
The inset in figure~\ref{fig10} (the left-hand panel) shows that finite-size effects 
on $U_c$ are small for all examined values of $q_U$ and thus these results can be
satisfactorily extrapolated to the thermodynamic limit $L\to \infty$.

 We also obtained similar results for the half-filled band case
$n=1$ (see figure~\ref{fig10}, the middle panel). However, in this case the critical interaction 
strength $U_c(q_U)$ exhibits a universal behaviour, i.e., it does not
depend on the value of the long-range hopping parameter $q_t$. Since in the
half-filled band case and $q_U=0$, the ground states of the ordinary 
($q_t=0$) as well as the generalized ($q_t >0$) Hubbard model are 
non-ferromagnetic~\cite{Fark1}, the existence of the ferromagnetic
state at this filling is purely a consequence of long-range interactions. 
Moreover, analysing the behaviour
of $U_c(q_U)$ for $q_U \to 0$, we found that only for $q_U=0$ 
(the Hubbard model with one-site interaction) $U_c=\infty$, while for finite
$q_U$ (that represents a much more realistic type of interactions between 
electrons on a lattice), the critical interaction strength $U_c$ is finite.
Thus, the absence of ferromagnetism in an ordinary model as well as in a generalized
Hubbard model with long-range hopping at $n \leqslant 1$ could be explained as 
a consequence of a too simplified description of electron-electron
interactions on the lattice. For any $q_U>0$, ferromagnetism comes naturally
from the Hubbard model with long-range interactions for a wide range of model
parameters without any other assumptions. As it is shown in the inset in
figure~\ref{fig10} (the middle panel),
our numerical results obtained for $n=1$ depend only very weakly on the 
size of the lattice and thus they can be satisfactorily extrapolated to 
macroscopic systems. 

Above the half-filled band case
$n>1$, we expect a strong interplay between the effects of long-range hopping 
and long-range interaction, since the long-range hopping itself stabilizes 
the ferromagnetic state for $n>1$~\cite{Fark1}, while the long-range 
interaction itself produces a fully opposite effect. The results of our 
numerical simulations obtained for $n=3/2$ are presented in figure~\ref{fig10}
(the right-hand panel) and they fully confirm this conjecture. Due to the
combined effects of the long-range hopping and the long-range Coulomb
interaction, the stability region of the ferromagnetic phase is reduced
in comparison to the on-site case ($q_U=0$), though it remains finite for
all the examined values of $q_t$. Similarly to the preceding cases, the 
results can be satisfactorily extrapolated to the thermodynamic limit
$L\to \infty$ and can be used for a description of macroscopic systems.

\subsection{The effect of flat bands}
\subsubsection{One-dimensional case}

In our previous paper we  showed~\cite{farky2} that the ferromagnetic state
can be also found in the so-called static limit of the Hubbard model,
where only one kind of electrons, say with spins up, can move on the
lattice, while electrons of the opposite spins are immobile. In this limit,
the ferromagnetic state is stabilized for a wide range of the on-site
Coulomb interaction $U$ between the up and down spin electrons for both
the hole doped case (the total concentration of electrons $n < 1$) and for 
the electron doped case ($n>1$). From this point of view, it is interesting to
ask whether the ferromagnetic state found for the zero value of the down-spin
electron hopping integral $t_{\downarrow}$ persists also at finite
$t_{\downarrow}$, or vanishes discontinuously as soon as $t_{\downarrow} > 0$.
To answer this question, we   numerically examined using the density matrix
renormalization group (DMRG) method~\cite{Peschel,Schollwock,Hallberg},
the asymmetric Hubbard model~\cite{AHM,AHM_a,AHM_b} 
($0 \leqslant t_{\downarrow} \leqslant 1$) that incorporates both the full Hubbard 
model $t_{\uparrow}=t_{\downarrow}$ and its static ($t_{\downarrow}=0$)
limit (in the rare-earth community also known  as the Falicov-Kimball
model~\cite{FKM}).

The Hamiltonian of the asymmetric Hubbard model is
\begin{equation}
H=-t_{\uparrow}\sum_{<ij>}c^+_{i\uparrow}c_{j\uparrow}
-t_{\downarrow}\sum_{<ij>}c^+_{i\downarrow}c_{j\downarrow}
+U\sum_{i}c^+_{i\uparrow}c_{i\uparrow}c^+_{i\downarrow}c_{i\downarrow},
\label{Eq8} 
\end{equation}
where $c^+_{i\uparrow}$ $(c_{i\uparrow})$ and $c^+_{i\downarrow}$
$(c_{i\downarrow})$
is the creation (annihilation) operator  of light ($\uparrow$) and heavy 
($\downarrow$) electron at the lattice site $i$.
  
The first two terms of (\ref{Eq8}) are the kinetic energies corresponding to
quantum-mechanical hopping of up-spin and down-spin electrons
between the nearest neighbor sites $i$ and $j$
with hopping probabilities $t_{\uparrow}$ and $t_{\downarrow}$, respectively.
The third term represents the on-site
Coulomb interaction between the up-spin electrons with density
$n_{\uparrow}=\frac{1}{L}\sum_id^+_{i\uparrow}d_{i\uparrow}$ and 
the down-spin electrons with density
$n_{\downarrow}=\frac{1}{L}\sum_id^+_{i\downarrow}d_{i\downarrow}$. 
The model is called ``asymmetric'' because the hopping integrals
for up-spin and down-spin electrons may be different. Usually, the hopping
integral of the up-spin electrons is taken to be the unit of energy
$(t_{\uparrow}=1)$ and the down-spin-electron hopping integral is considered 
in the limit $t_{\downarrow} \leqslant 1$. This is the reason why the up-spin electrons 
are called light ones and the down-spin electrons are called heavy. 

Based on the  results obtained in the static limit of the model~\cite{farky2}, 
 we  chose three different concentration
limits, and namely, $n=1/4, 1/2$ and $3/4$, at which we  performed
exhaustive numerical studies of the asymmetric Hubbard model, with a goal
to reveal the effects of the down-spin electron hopping on the stability
of the ferromagnetic state found at $t_\downarrow=0$. In all three limits,
the comprehensive magnetic phase diagrams of the model in the 
$t_\downarrow$--$U$ plane are constructed using the DMRG method, that allows us
to treat relatively large clusters with high accuracy.  To identify different
magnetic phases that enter  the ground-state phase diagrams, we calculated the DMRG ground-state energy $E_{\text{DMRG}}$, the total spin $S$ in the 
DMRG ground state by evaluating 
\begin{eqnarray}
\langle \bf{S}^2\rangle = \sum_{i,j}\langle \bf{S_i}\bf{S_j}\rangle
\label{Eq9}
\end{eqnarray}
and the energy of the fully polarized state $E_f$, which is exactly known
since the state has no double occupancy. Different phases are then
classified as follows: (i) the fully polarized state, $E_{\text{DMRG}}>E_f$ and
$S=S_{\text{max}}$, (ii) the partially polarized state,  $E_{\text{DMRG}}<E_f$ and $S\neq S_{\text{max}}$ and (iii) the non-polarized state, $E_{\text{DMRG}}<E_f$ and $S=0$.

\begin{figure}[h!]
\begin{center}
\includegraphics[width=10cm]{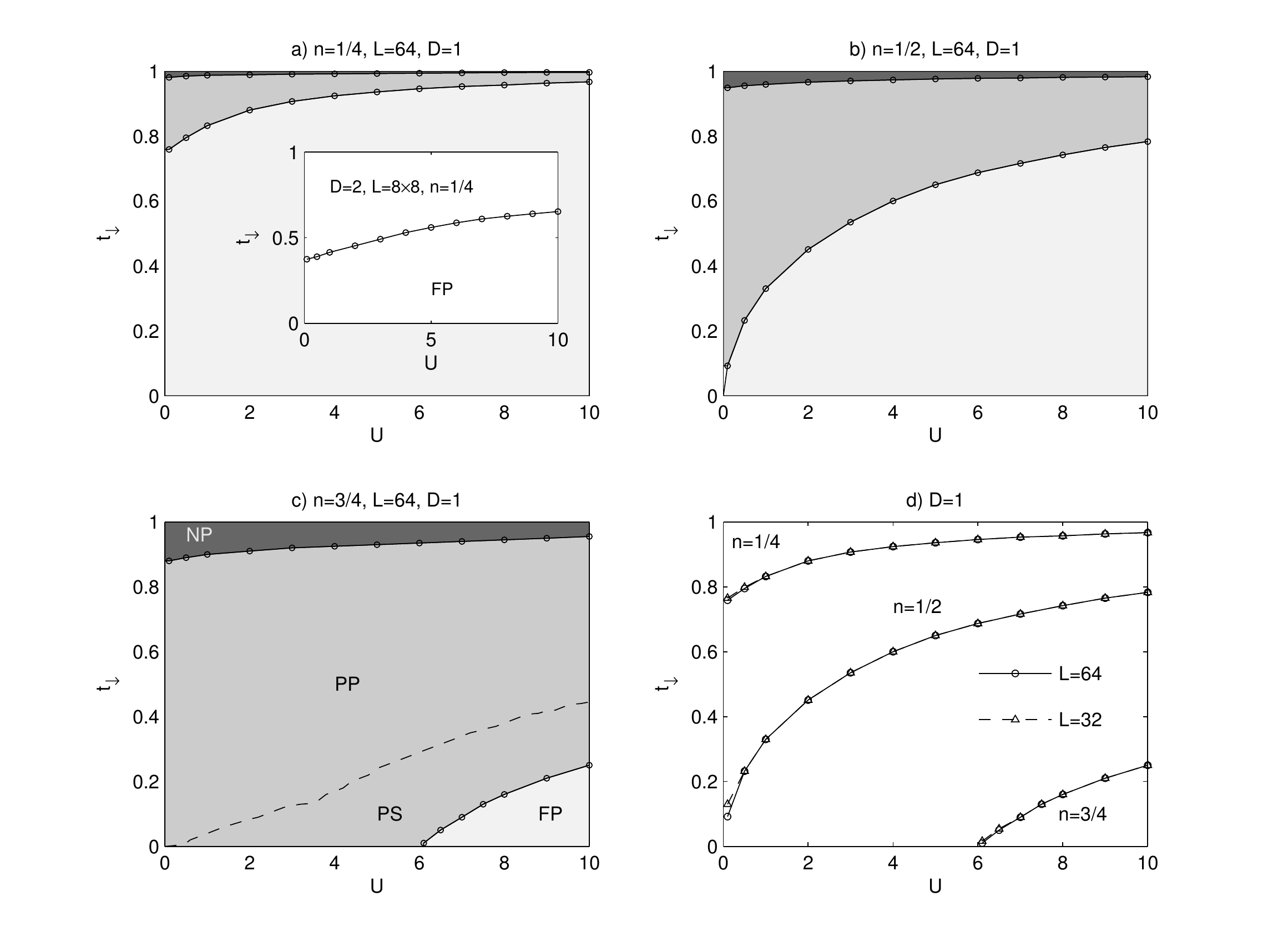}
\end{center}
\caption{Phase diagrams of the asymmetric Hubbard model in $D=1$ calculated
for three electron concentrations ($n=1/4, n=1/2$ and $n=3/4$).
Different phases correspond to the non-polarized NP phase (the black areas),
the partially polarized PP phase (the gray areas) and the fully polarized FP
phase (the light gray areas). In the PS region ($n=3/4$), the ground states
are phase separated. The inset in figure~\ref{fig11}a shows the results for $D=2$ obtained
on $L=8\times8$ cluster. Figure~\ref{fig1}d presents finite-size effects on the phase
boundary between the FP and PP phase in $D=1$~\cite{Fark5}.}
\label{fig11}
\end{figure}

The resultant magnetic phase diagrams are presented in figure~\ref{fig11}. For small
electron concentrations ($n=1/4$), the ferromagnetic domain takes the largest
part of the phase diagram. The ferromagnetic state is stable for all
values of the down-spin electron hopping integrals $t_\downarrow > 0.75$,
and its stability region is further stabilized with an increasing Coulomb
interaction $U$. The non-polarized state is stable only in a very narrow
region near the full Hubbard limit ($t_\uparrow=t_\downarrow=1$) and
thereby it becomes extremely sensitive to small changes in
$t_\uparrow$ and $t_\downarrow$. This is probably also the reason why the
ferromagnetic state can be stabilized by the correlated hopping~\cite{Amadon},
when the original electron hopping amplitudes are reduced according
to whether the sites $i$ and $j$ are occupied or are not occupied by the electron
of the opposite spin. The qualitatively same picture is also observed  for
intermediate electron concentrations ($n=1/2$), with only one exception,
and namely, that the ferromagnetic phase is now reduced against the non-polarized and partially polarized phases. This trend also holds  for higher
electron concentrations ($n=3/4$), when the ferromagnetic phase persists
only in the strong coupling limit, exactly in accordance with conclusions
obtained for the static limit $t_\downarrow=0$. In addition, using two
complementary methods based on the calculations of the structure
factors~\cite{Gu} and the most probable distribution of heavy
particles~\cite{farky3} we  found that with an exception of the partially
polarized phase (at $n=3/4$), all remaining phases are homogeneous. 
The partially polarized phase found at $n=3/4$ is homogeneous only for
 sufficiently large $t_{\downarrow}$ while in the opposite limit, the phase
separation takes place. This result accords with the ones obtained
for the case of the asymmetric Hubbard model in which both $n_{\uparrow}$
and $n_{\downarrow}$ are fixed~\cite{Gu,farky3,Ueltschi}.

To exclude the influence of finite-size effects on the magnetic-phase
diagrams of the asymmetric Hubbard model, we  also calculated the phase
boundaries for different finite clusters of $L=32$ and $L=64$ sites.
The results of numerical calculations obtained for $n=1/4,1/2$ and $3/4$
are shown in figure~\ref{fig11}d and they clearly show that finite size effects are
negligible and thus the results obtained can be satisfactorily extrapolated
to the macroscopic systems.

\subsubsection{Two-dimensional case}

From the point of view of real materials, the question of fundamental
importance is whether these results also persist  in higher dimensions. For
this reason, we also  performed similar calculations  in two
dimensions.  Since the DMRG method does not work very well in $D>1$,
especially for the calculation of long-range correlation functions 
like $\langle \bf{S_i}\bf{S_j}\rangle$,  instead of the DMRG
method we  used the projector quantum Monte Carlo method~\cite{Sorella,Loh,Imada}.             
The quantum Monte Carlo simulations were performed using a projector
algorithm which applies $\exp(-\theta H)$ to a trial wave-function (in our
case, the solution for $U = 0$). A projector parameter $\theta \sim 30$
suffices to reach well converged values of the observables discussed here.
A time slice of $\Delta \theta = 0.05$ was used in general.
The resultant numerical solutions for the critical values of the down-spin
electron hopping integrals $t_\downarrow$ below which the ground state
is ferromagnetic are shown in the inset in figure~\ref{fig11}a. One can see that
the ferromagnetic state is also robust  in two dimensions although
the corresponding values of $t_\downarrow$ in two dimensions are slightly
smaller than the ones in $D=1$.

\subsection{Influence of lattice structure}
\subsubsection{Short-range hopping}

\begin{figure}[!t]
\begin{center}
\hspace{-0.5cm}
\includegraphics[width=5.7cm]{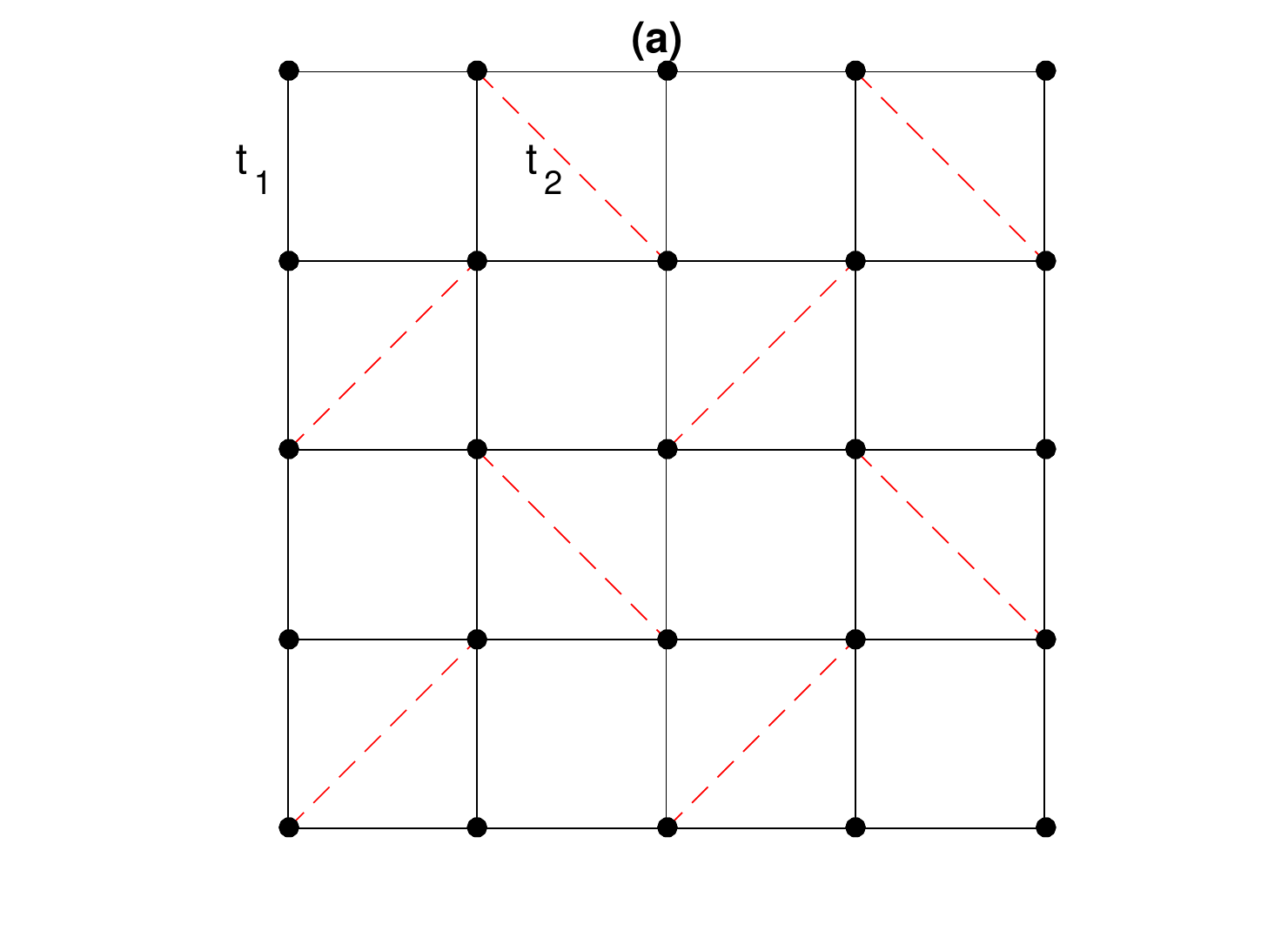}
\hspace{-1.3cm}
\includegraphics[width=5.7cm]{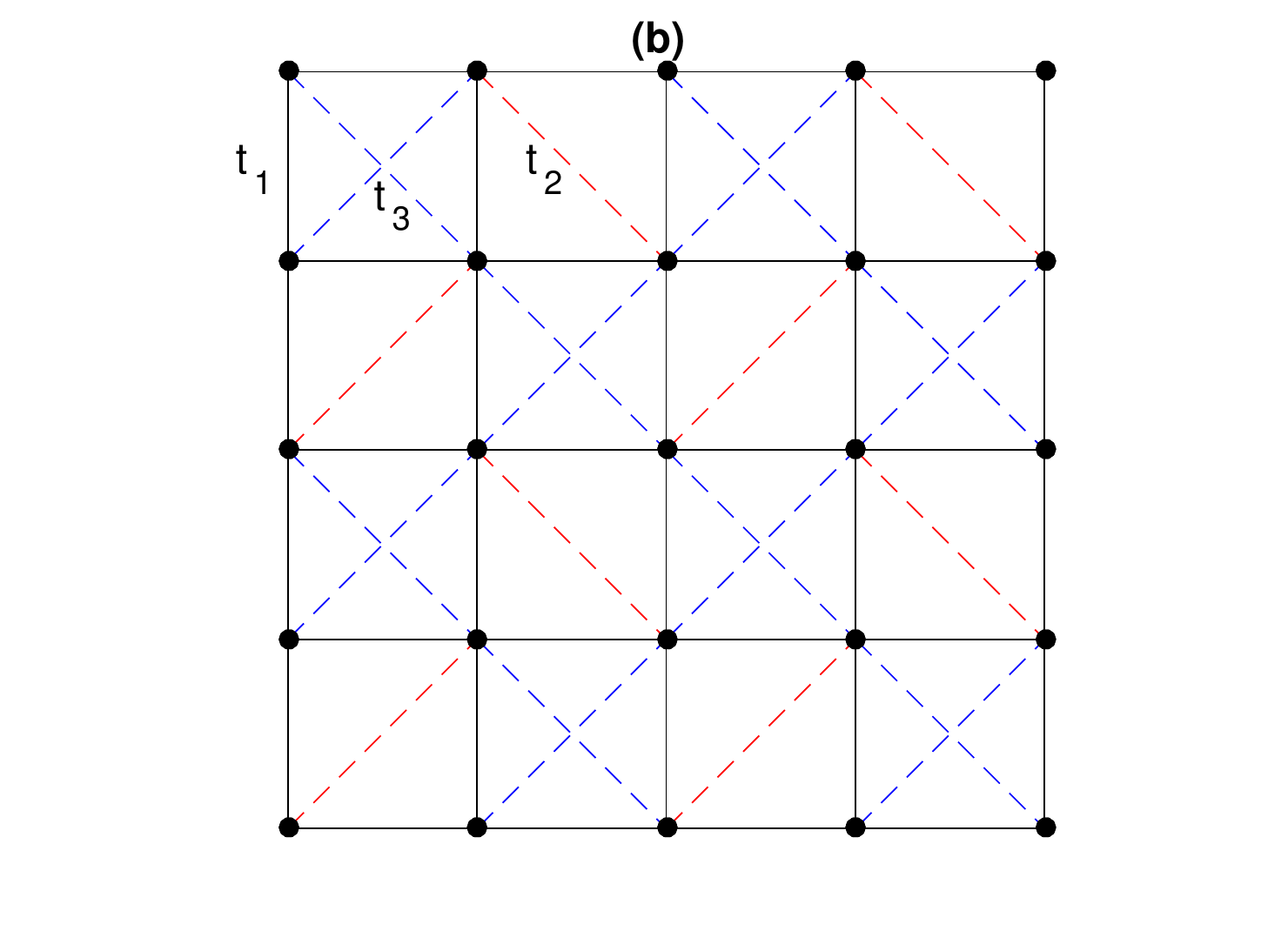}
\hspace{-1.3cm}
\includegraphics[width=5.7cm]{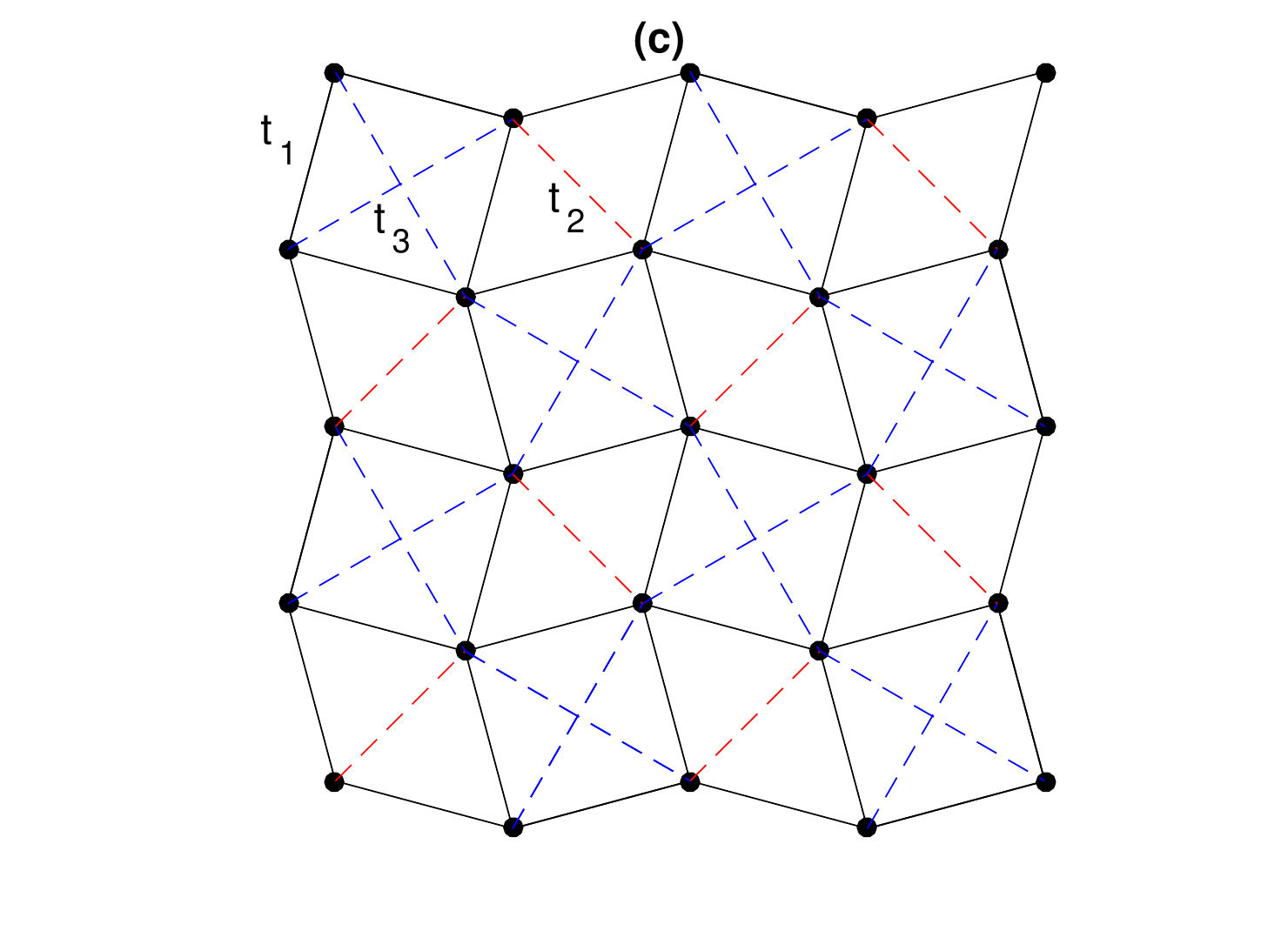}
\end{center}
\caption{(Colour online) (a) The original SSL with the first ($t_1$) and second ($t_2$)
nearest-neighbor couplings. (b) The generalized SSL with the first ($t_1$), 
second ($t_2$) and third ($t_3$) nearest-neighbor couplings, and (c)
the topologically identical structure realized in the (001) plane of
rare-earth tetraborides~\cite{Fark6}.   
}
\label{fig12}
\end{figure}

The results presented in the previous sections indicate that the lattice 
structure, which dictates the shape of the DOS, plays an important role in 
stabilizing the ferromagnetic state.  
Motivated by these results, we  performed the same calculations 
on the special type of lattice, the so-called Shastry-Sutherland 
lattice (SSL). The SSL represents one of the simplest systems with 
geometrical frustration, so that putting the electrons on this lattice, 
one can simultaneously examine  both the effect of interaction and 
the effect of geometrical frustration on the ground-state properties of the 
Hubbard model. This lattice was first introduced by Shastry 
and Sutherland~\cite{Shastry} as an interesting example of 
a frustrated quantum spin system with an exact ground state. 
It can be described as a square lattice with 
the nearest-neighbor links $t_1$ and the next-nearest neighbors
links $t_2$ in every second square (see figure~\ref{fig12}a).
The SSL attracted much attention after its experimental realization
in the SrCu$_2$(BO$_3$)$_2$ compound~\cite{Kageyama2}. The observation
of a fascinating sequence of magnetization plateaus (at 
$m/m_s=$1/2, 1/3, 1/4 and 1/8 of  the  saturated  magnetization $m_s$) 
in this material~\cite{Kodama} stimulated further theoretical and experimental 
studies of the SSL. 
The SSL with the first, second and third nearest-neighbor links is 
shown in figure~\ref{fig12}b and this is just the lattice that will be used in our 
next numerical calculations.    

Thus, our starting Hamiltonian, corresponding to the one band Hubbard
model on the SSL, can be written as follows:
\begin{equation}
H=
-t_1\sum_{\langle ij \rangle_1, \sigma}c^+_{i\sigma}c_{j\sigma}
-t_2\sum_{\langle ij \rangle_2, \sigma}c^+_{i\sigma}c_{j\sigma}
-t_3\sum_{\langle ij \rangle_3, \sigma}c^+_{i\sigma}c_{j\sigma}
+U\sum_{i}n_{i\uparrow}n_{i\downarrow}.
\label{Eq10}
\end{equation}
The first three terms of (1) are the kinetic energies
corresponding to the quantum-mechanical hopping of electrons
between the first, second and third nearest neighbors and the
last term is the Hubbard on-site repulsion between two electrons
with opposite spins. We set $t_1=1$ as the energy unit and thus
$t_2$ ($t_3$) can be seen as a measure of the frustration strength.   
It should be noted that most of the papers on the SSL concern 
various spin models, while there are only a few papers concerning
the interacting electrons (the Hubbard model) on the SSL~\cite{Liu}.

\begin{figure}[!t]
\begin{center}
\includegraphics[width=7cm]{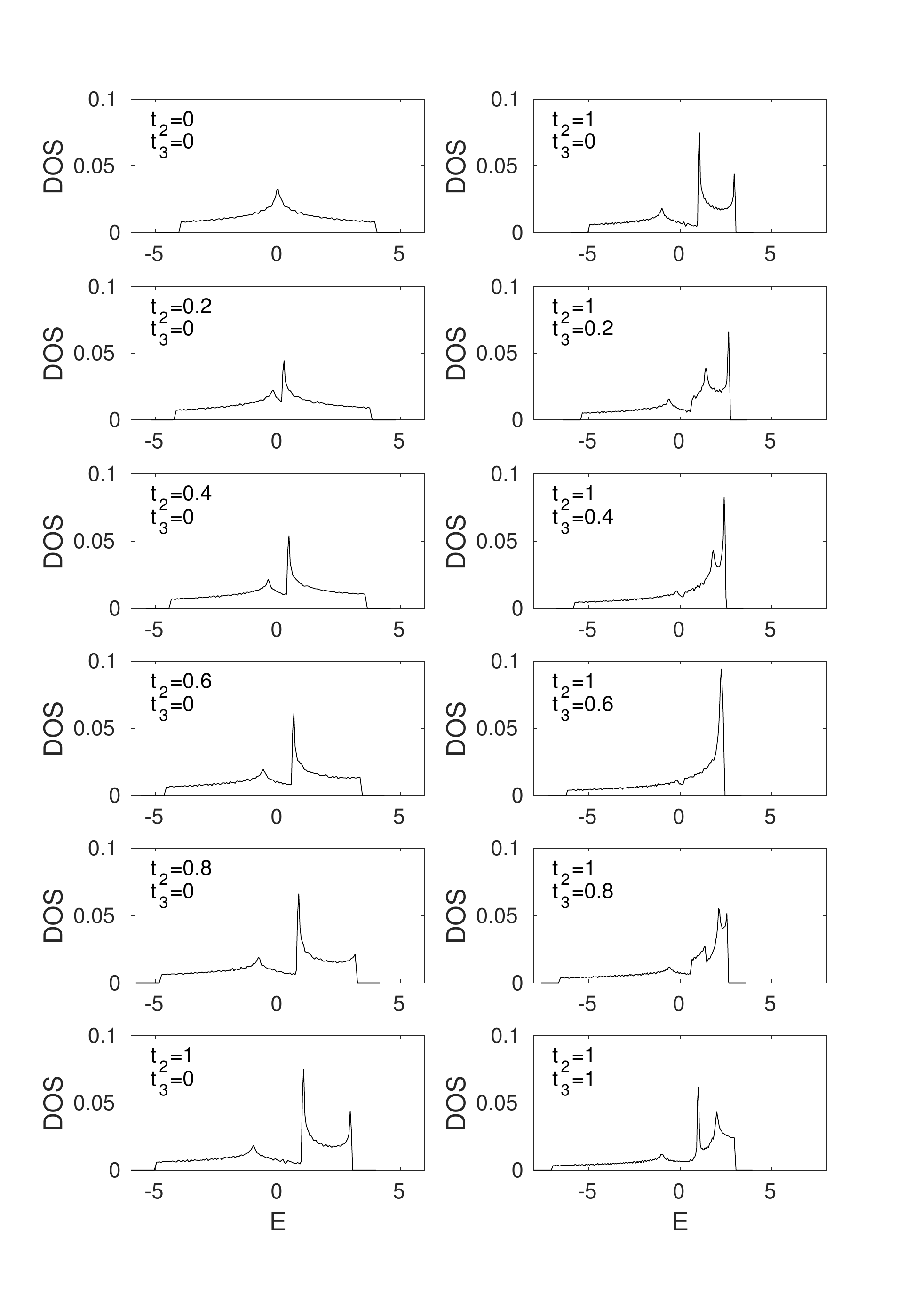}
\end{center}
\caption{ Non-interacting DOS calculated numerically for different 
values of $t_2$ and $t_3$ on the finite cluster of $L=200 \times 200$ sites~\cite{Fark6}.}
\label{fig13}
\end{figure}

\begin{figure}[!t]
\begin{center}
\includegraphics[width=6cm]{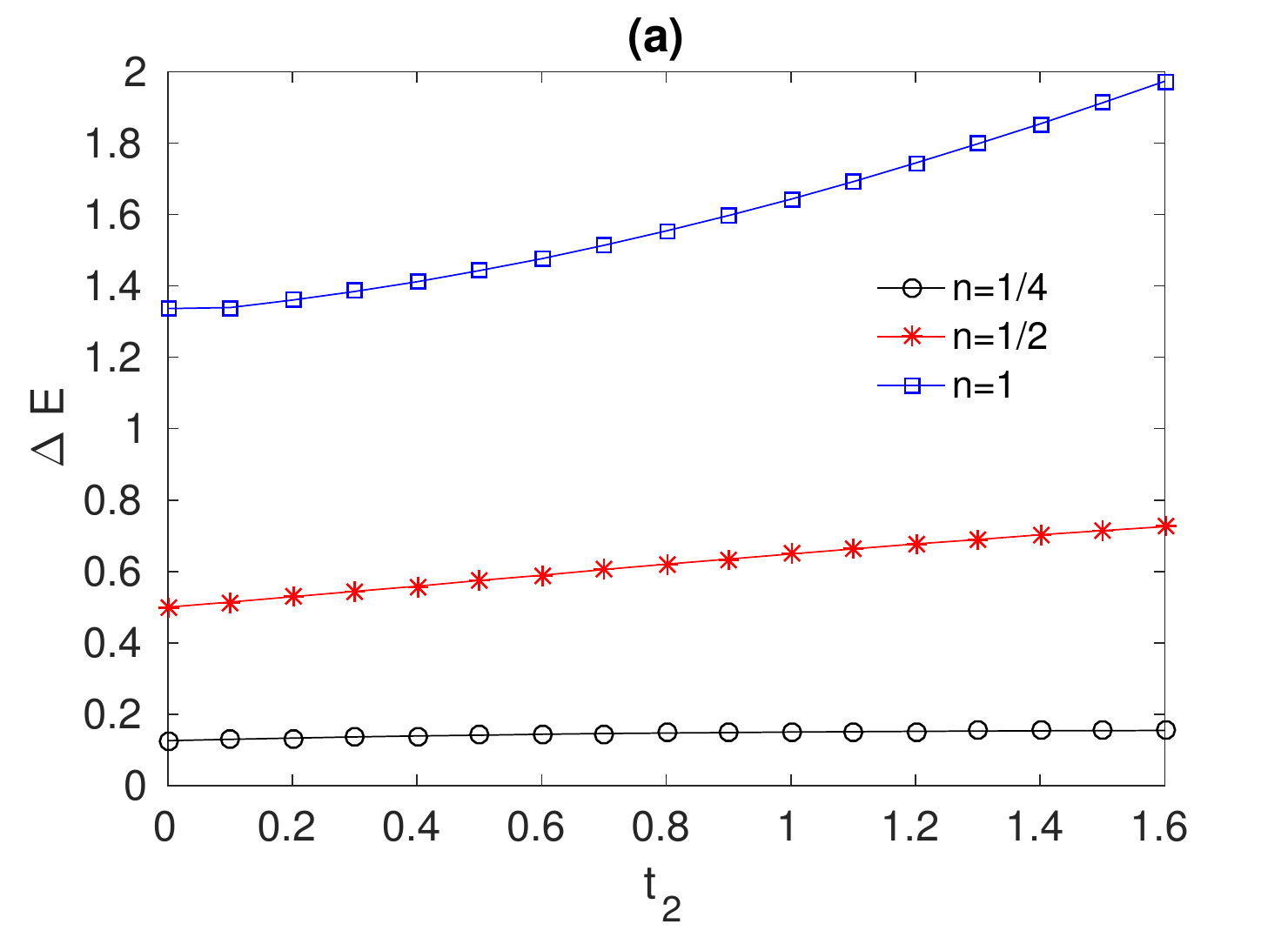}
\includegraphics[width=6cm]{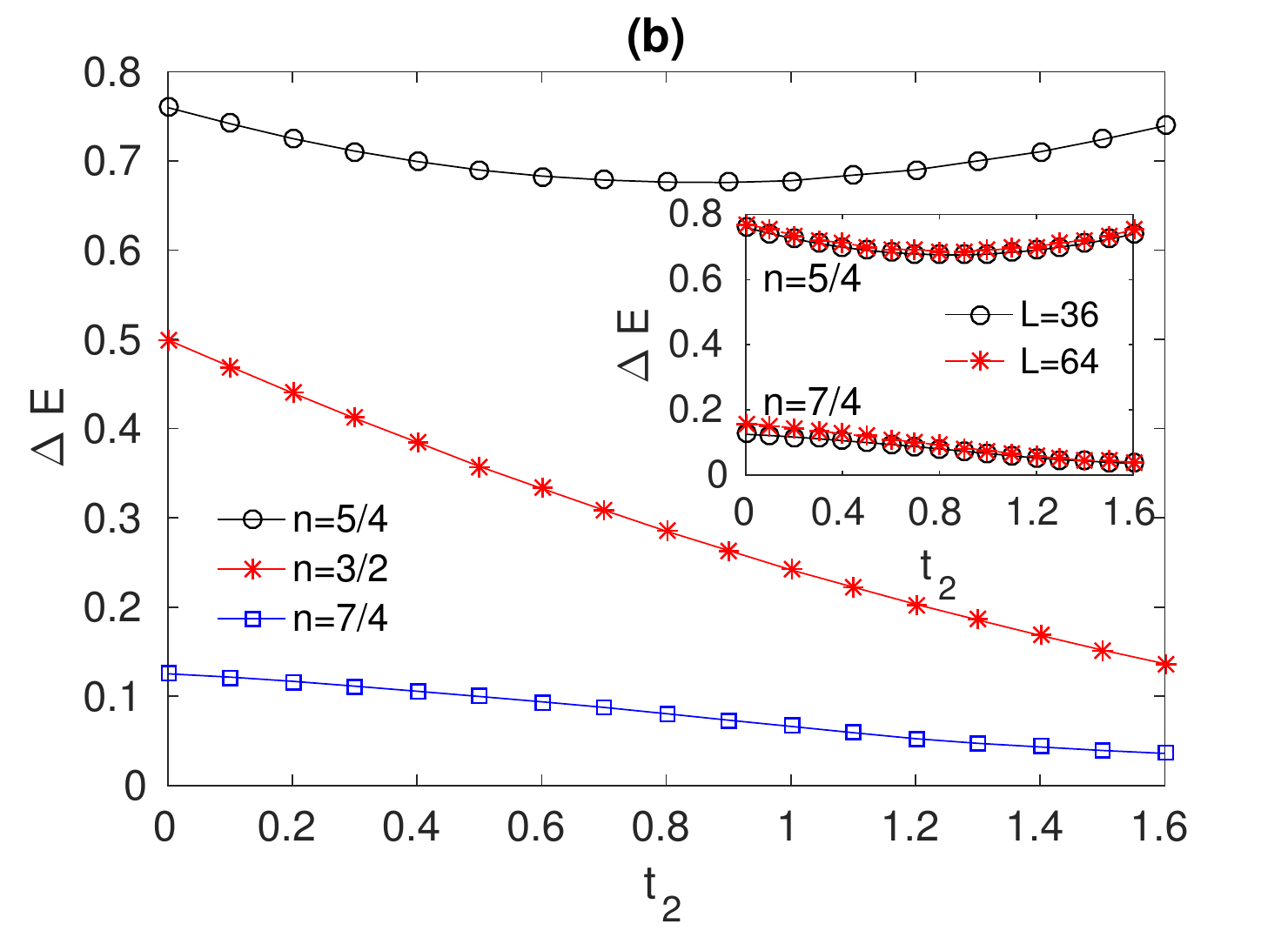}
\end{center}
\caption{(Colour online)  The difference $\Delta E=E_f-E_{\text{min}}$ between the 
ferromagnetic state $E_f$ and the lowest ground-state energy 
from $E_g(S_z)$ as a function of the frustration parameter $t_2$
calculated for $n\leqslant 1$ (a) and $n>1$ (b) on the finite cluster 
of $L=6 \times 6$ sites ($U=1, t_3=0$). The inset shows $\Delta E$, 
calculated for two different electron densities on clusters of 
$L=6 \times 6$ and $L=8 \times 8$ sites~\cite{Fark6}.}
\label{fig14}
\end{figure}

To identify the nature of the ground state of the Hubbard model 
on the SSL we have used the small-cluster-exact-diagonalization 
(Lanczos) method~\cite{Lanczos} and the projector quantum Monte Carlo 
method~\cite{Imada}. 
In both cases, the numerical calculations proceed in the following steps. 
Firstly, the ground-state energy of the model $E_g(S_z)$ is 
calculated in all different spin sectors $S_z=N_{\uparrow}-N_{\downarrow}$ 
as a function of the model parameters $t_2,t_3$ and $U$. Then, the resulting
behaviors of $E_g(S_z)$  are used directly to identify the regions
in the parametric space of the model, where the fully polarized state 
has the lowest energy.
To reveal the possible stability regions of the ferromagnetic
state in the Hubbard model on the SSL, let us first examine
the effects of the geometrical frustration, represented by nonzero
values of $t_2$ and $t_3$, on the behavior of the non-interacting DOS. 
The previous numerical studies of the standard one-dimentional and two-dimensional 
Hubbard model with next-nearest~\cite{M_H} as well as 
long-range~\cite{Fark1,Fark2} hopping showed that just this quantity 
could be used as a good indicator for the emergence of ferromagnetism
in the interacting systems. 
The noninteracting DOS of the $U=0$ Hubbard model on the SSL 
of size $L=200 \times 200$, obtained by exact diagonalization 
of $H$ (for $U=0$) is shown in figure~\ref{fig13}. The left-hand panels correspond 
to the situation when $t_2 >0$ and $t_3=0$, while the right-hand panels correspond 
to the situation when both $t_2$ and $t_3$ are finite.
Numerical calculations were performed for different values 
of the delta function broadening $\epsilon$ and it was found that 
the value of $\epsilon=0.02$ (the case presented in this section) 
is sufficient to satisfactorily reproduce  all significant features of DOS.
One can see that once the frustration parameter $t_2$ is nonzero,
the spectral weight starts to shift to the upper band edge and
the noninteracting DOS becomes strongly asymmetric. Thus, taking
into account the above mentioned scenario, there is a real chance
that the interacting system could be ferromagnetic in the limit
of high electron concentrations. To verify this conjecture, we  
performed exhaustive numerical studies of the model Hamiltonian~(\ref{Eq10})
for a wide range of the model parameters $U,t_2$ and $n$ at $t_3=0$. 
Typical results of our PQMC calculations obtained on finite cluster of 
$L=6\times6$ sites, in two different concentration limits 
($n\leqslant 1$ and $n>1$) are shown in figure~\ref{fig14}. 
There is plotted the difference $\Delta E=E_f-E_{\text{min}}$ between 
the fully saturated ferromagnetic state $E_f$, which can be calculated 
exactly and the lowest ground-state energy from $E_g(S_z)$ as a function 
of the frustration parameter $t_2$. According to this definition, 
the ferromagnetic state, strictly referred to as the fully saturated ferromagnetic 
state, corresponds to $\Delta E=0$ (everywhere in the paper, the notation 
ferromagnetism (ferromagnetic state) concerns the fully saturated 
ferromagnetism (the fully saturated ferromagnetic state)). It is seen that
for electron concentrations below the half filled band case $n=1$,
$\Delta E$ is the increasing function of $t_2$, and thus
there is no sign  of stabilization of the ferromagnetic state
for $n\leqslant 1$, in accordance with the above mentioned 
scenario. 

The situation looks more promising in the opposite 
limit $n>1$. In this case, $\Delta E$ is considerably reduced
with an increasing $t_2$, though this reduction is still insufficient
to reach the ferromagnetic state $\Delta E=0$ for physically 
reasonable values of $t_2$ ($t_2<1.6$) that correspond to the situation
in the real materials. To exclude the finite-size effect, we  also 
performed the same calculations on the larger cluster
of $L=8 \times 8$ sites, but again no signs of stabilization of the
ferromagnetic state were observed (see the inset to figure~\ref{fig14}b).

\begin{figure}[!t]
\begin{center}
\includegraphics[width=6cm]{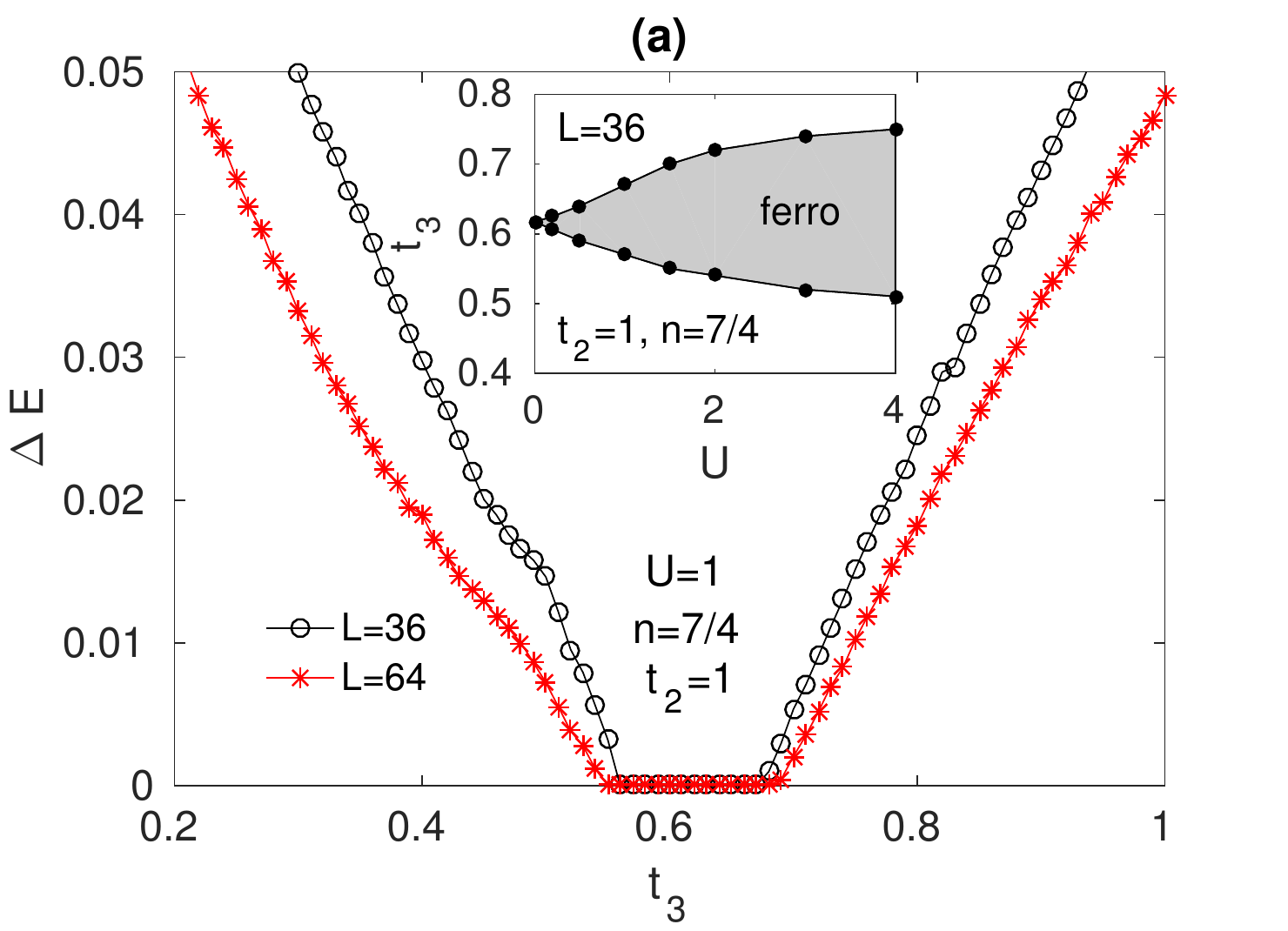}
\includegraphics[width=6cm]{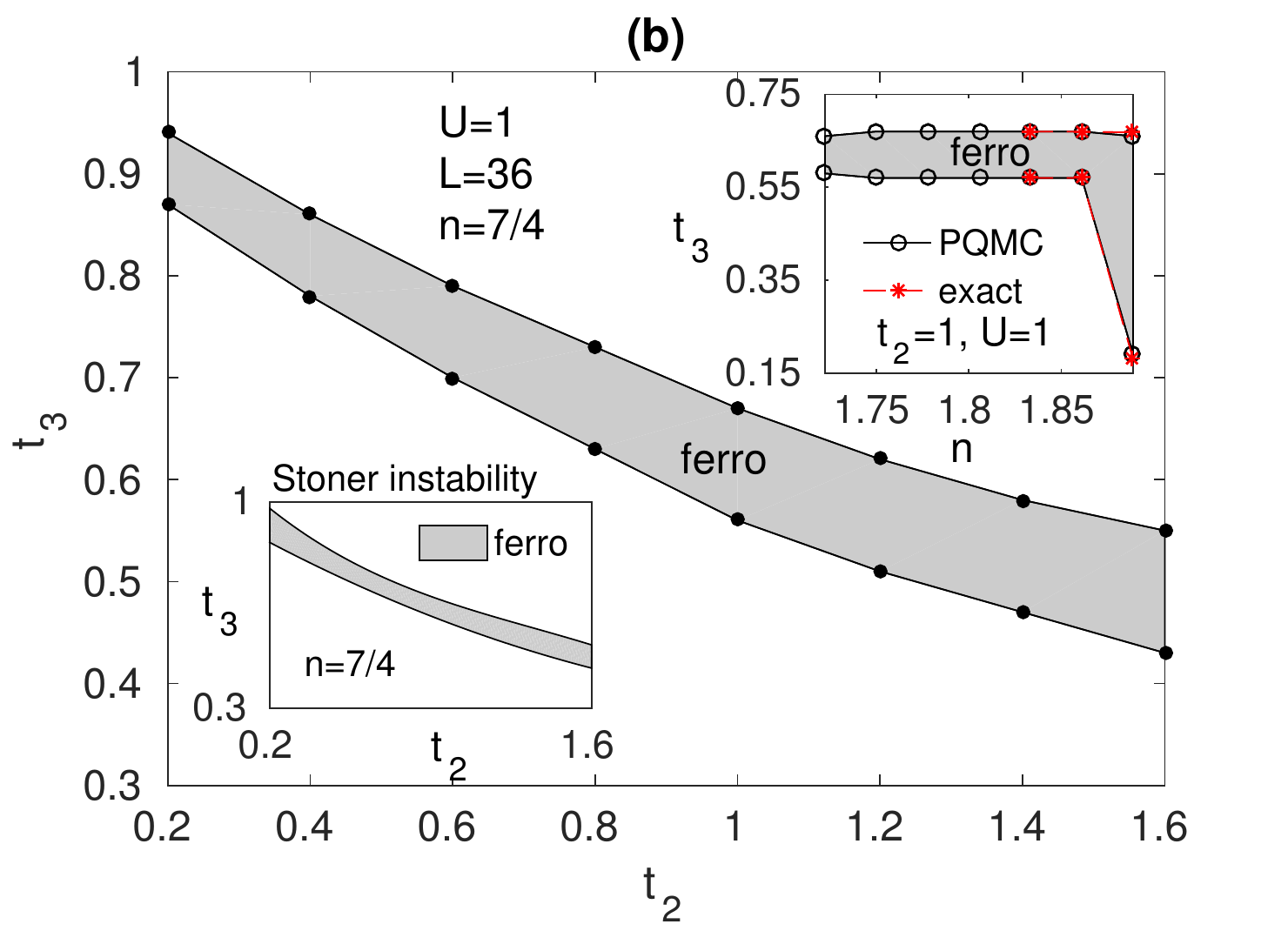}
\end{center}
\caption{(Colour online) (a) The difference $\Delta E=E_f-E_{\text{min}}$ as a function of 
the frustration parameter $t_3$ calculated for $U=1, t_2=1, n=7/4$ and two 
different finite clusters of $L=6 \times 6$ and $L=8 \times 8$ sites. The
inset shows the ground-state diagram of the model in the $t_3$--$U$ plane.
(b) The comprehensive phase diagrams of the model in the $t_3$--$t_2$
and $t_3$--$n$ plane~\cite{Fark6}.}
\label{fig15}
\end{figure}

For this reason, we  turned our attention to
the case $t_2>0$ and $t_3>0$. The noninteracting DOS corresponding
to this case is displayed in figure~\ref{fig13} (the right-hand panels). These
panels clearly demonstrate that with the increasing value of the 
frustration parameter $t_3$, a more spectral weight is still shifted 
to the upper band edge. A special situation arises at $t_3=0.6$,
when the spectral weight is strongly peaked at the upper band edge.
In this case, the nonintercting DOS is practically identical to the 
one corresponding to noninteracting electrons with long-range
hopping~\cite{Fark1,Fark2}.
Since the long-range hopping supports ferromagnetism in the standard
Hubbard model for electron concentrations above the half-filled band 
case~\cite{Fark1,Fark2}, we expect that this could be  also true 
for the Hubbard model on the SSL, at least for some values of frustration 
parameters $t_2$ and $t_3$. Therefore, we  decided to perform numerical 
studies of the model for a wide range of $t_3$ values at fixed $t_2, U$ 
and $n$ ($t_2=1, U=1, n=7/4$). To minimize the finite-size effects,
the  numerical calculations were done on two different finite clusters 
of $L=6\times 6$ and $L=8\times 8$ sites. The results of our calculations
for $\Delta E$ as a function of $t_3$ are displayed in figure~\ref{fig15}a. 
In accordance with the above mentioned assumptions, we find a relatively 
wide region of $t_3$ values around $t_3=0.6$, where the ferromagnetic 
state is stable. It is seen that the finite-size effects on the stability
region of the ferromagnetic phase are negligible and thus these
results can be satisfactorily extrapolated to the thermodynamic limit
$L=\to \infty$. Moreover, the same calculations performed 
for different values of the Hubbard interaction $U$ showed that the correlation 
effects  (nonzero $U$) further stabilize the ferromagnetic state and lead 
to the emergence of macroscopic ferromagnetic domain in the $t_3$--$U$ 
phase diagram (see inset to figure~\ref{fig15}a). This confirms the crucial role
of the Hubbard interaction $U$ in the mechanism of stabilization 
of ferromagnetism on the geometrically frustrated lattice.
In figure~\ref{fig15}b, we plotted the comprehensive
phase diagrams of the model in the $t_3$--$n$ as well as $t_3$--$t_2$ plane, 
which clearly demonstrate that the ferromagnetic state is robust 
with respect to doping  and frustration. 
In addition, we  also calculated the stability region of the 
ferromagnetic state using the well-known Stoner criterion, and
surprisingly, we  found a nice correspondence of the results over 
the whole region of $t_2$ and $t_3$ values (see the lower inset 
to figure~\ref{fig15}b), despite the fact that the Stoner criterion 
is in general a crude approximation.

To check the convergence of PQMC results, we  performed the same 
calculations by the Lanczos exact diagonalization method. Of course, 
on such a large cluster, consisting of $L=6 \times 6$ sites, we were 
able to examine (due to high memory requirements) only several electron 
fillings near the fully occupied band ($N=2L$). The exact diagonalization 
and PQMC results for the width of the ferromagnetic phase 
obtained on finite cluster of $L=6 \times 6$ sites, for three
different electron fillings from the high concentration limit 
($N=66,67,68$), are displayed in the inset to figure~\ref{fig15}b and 
they show a nice convergence of PQMC results.    

\begin{figure}[!t]
\begin{center}
\includegraphics[width=8cm]{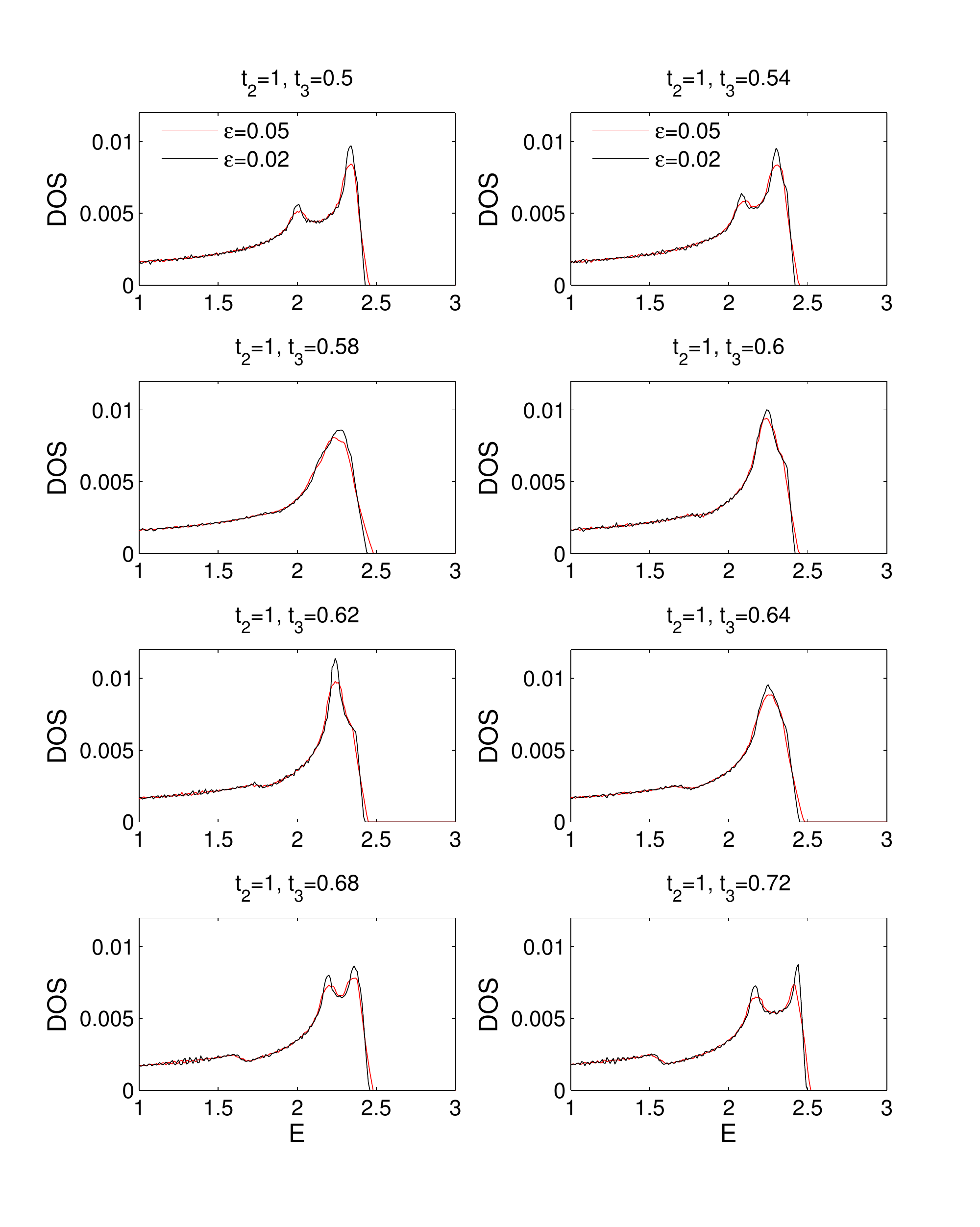}
\end{center}
\caption{(Colour online) Non-interacting DOS calculated numerically for $t_2=1$ 
and different values of~$t_3$ (near $t_3=0.6$) on the finite cluster 
of $L=200 \times 200$ sites~\cite{Fark6}.}
\label{fig16}
\end{figure}

Let us finally turn our attention to the question of possible connection
between ferromagnetism and the noninteracting DOS that is discussed
at the beginning of the paper. Figure~\ref{fig15}a and figure~\ref{fig15}b  show 
that for each finite $U$ and $n$ sufficiently large ($n \sim 7/4$), there exists a finite interval 
of $t_3$ values, around $t_3\sim 0.6$, where the ferromagnetic state is 
the ground state of the model. To examine a possible connection 
between ferromagnetism and the noninteracting DOS, we   numerically
calculated  the noniteracting DOS for several different 
values of $t_3$ from this interval and its vicinity. The results
obtained for $U=1, n=7/4$ and  $t_2=1$ are displayed in figure~\ref{fig16}.      
Comparing these results with the ones presented in figure~\ref{fig15}a for 
the stability region of the ferromagnetic phase at the same values
of $U,n$ and $t_3$, one can see that there is an obvious correlation 
between the shape of the noninteracting DOS and ferromagnetism. 
Indeed, the ferromagnetic state is stabilized only for these values 
of frustration parameters $t_2,t_3$, which lead to the single peaked 
noninterating DOS at the band edge. From the moment that two or more peaks appear 
in the noninteracting DOS at the band edge (by changing $t_2$ or $t_3$), 
ferromagnetism is suppressed.

\subsubsection{Long-range hopping}

\begin{figure}[!t]
\begin{center}
\includegraphics[width=8.0cm]{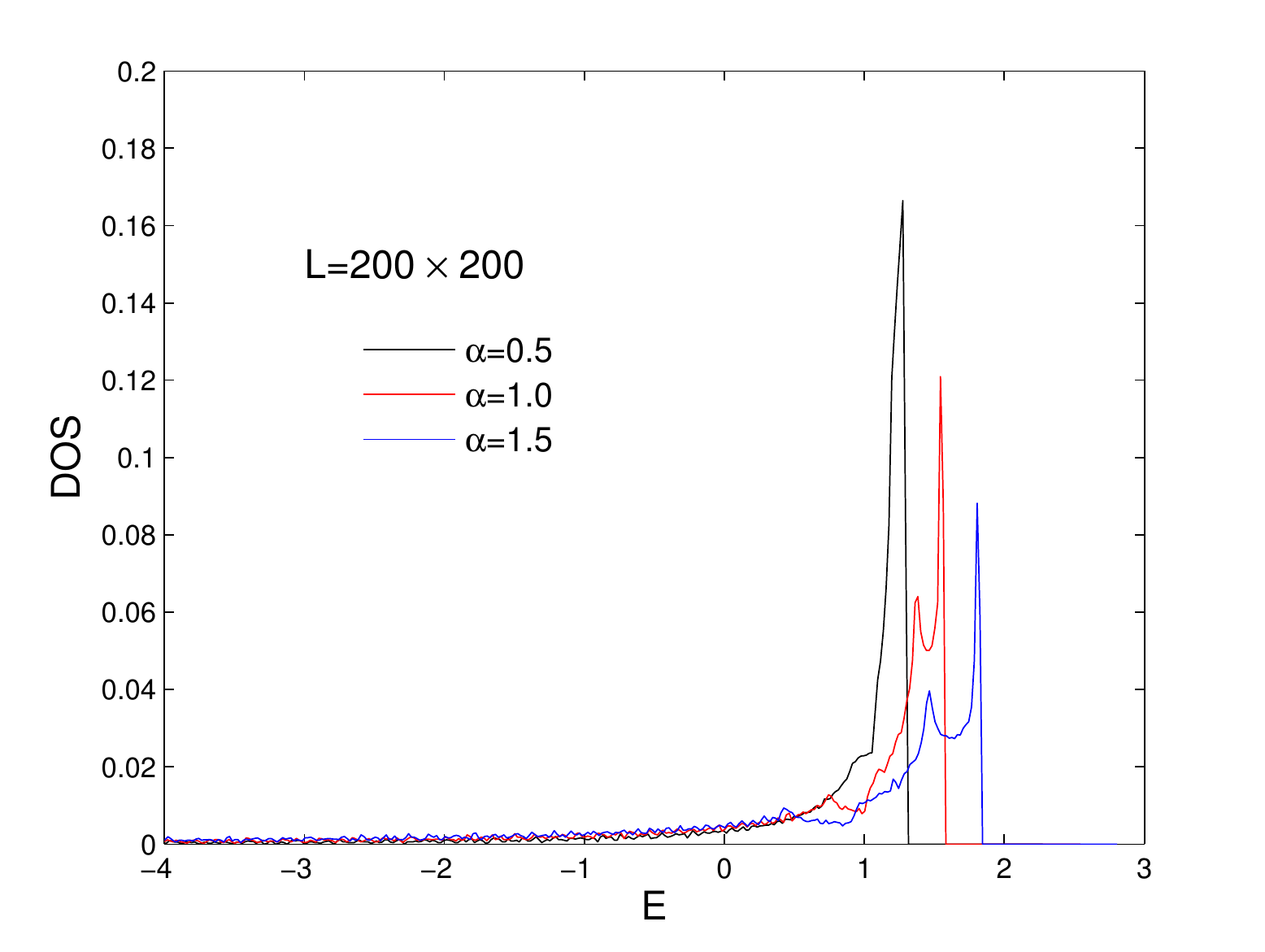}
\end{center}
\caption{(Colour online) Non-interacting DOS calculated numerically for different
values of the long-range hopping parameter $\alpha$ on the finite cluster 
of $L=200 \times 200$ sites~\cite{Fark7}.}
\label{fig17}
\end{figure}

With respect to the above presented results, it is natural to ask what happens 
in the situation when also  the next-nearest neighbor hopping terms will be included 
(for example, the $t_4$ and $t_5$ terms are of the same order as the $t_3$ term). 
To answer this question, we  performed the same calculations 
with the same one-parametric formula [equation~(\ref{Eq2})] as was used in the 
section 2.1. 
To reveal the possible stability regions of the ferromagnetic
state in the generalized Hubbard model on the SSL, let us first examine
the effects of the long-range hopping on the behavior of the non-interacting 
DOS. As mentioned above, just this quantity,  particularly, 
the appearance of the single-peaked DOS near the band edge 
could be used as a good indicator for the emergence of ferromagnetism
in the interacting systems. The noninteracting DOS of the $U=0$ Hubbard 
model on the SSL of size $L=200 \times 200$, obtained by exact diagonalization
of $H$ (for $U=0$) is shown in figure~\ref{fig17} for several different values ot the 
long-range-hopping parameter $\alpha=\ln q$. 
One can see that the DOS is 
strongly asymmetric with practically all spectral weight located at the upper
band edge, which in accordance with some previous
works indicates 
a possible region of ferromagnetism for electron concentrations above
the half-filed band case $n>1$. Moreover, the DOS is double-peaked for $\alpha \geqslant 1$ 
and single-peaked for $\alpha < 1$, indicating ferromagnetism in the limit 
of small values of $\alpha$.

\begin{figure}[!t]
\begin{center}
\includegraphics[width=7.5cm]{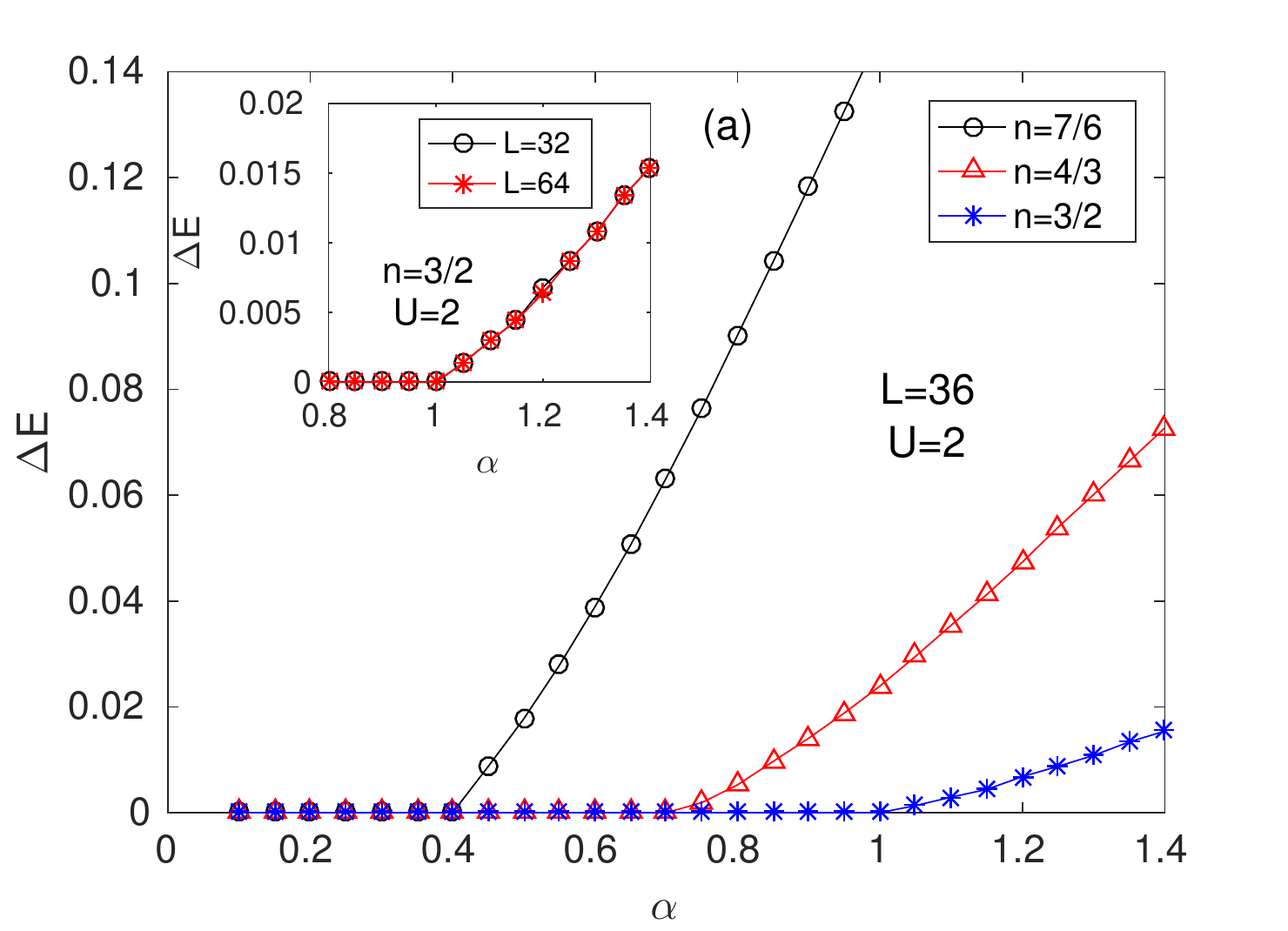}
\includegraphics[width=7.5cm]{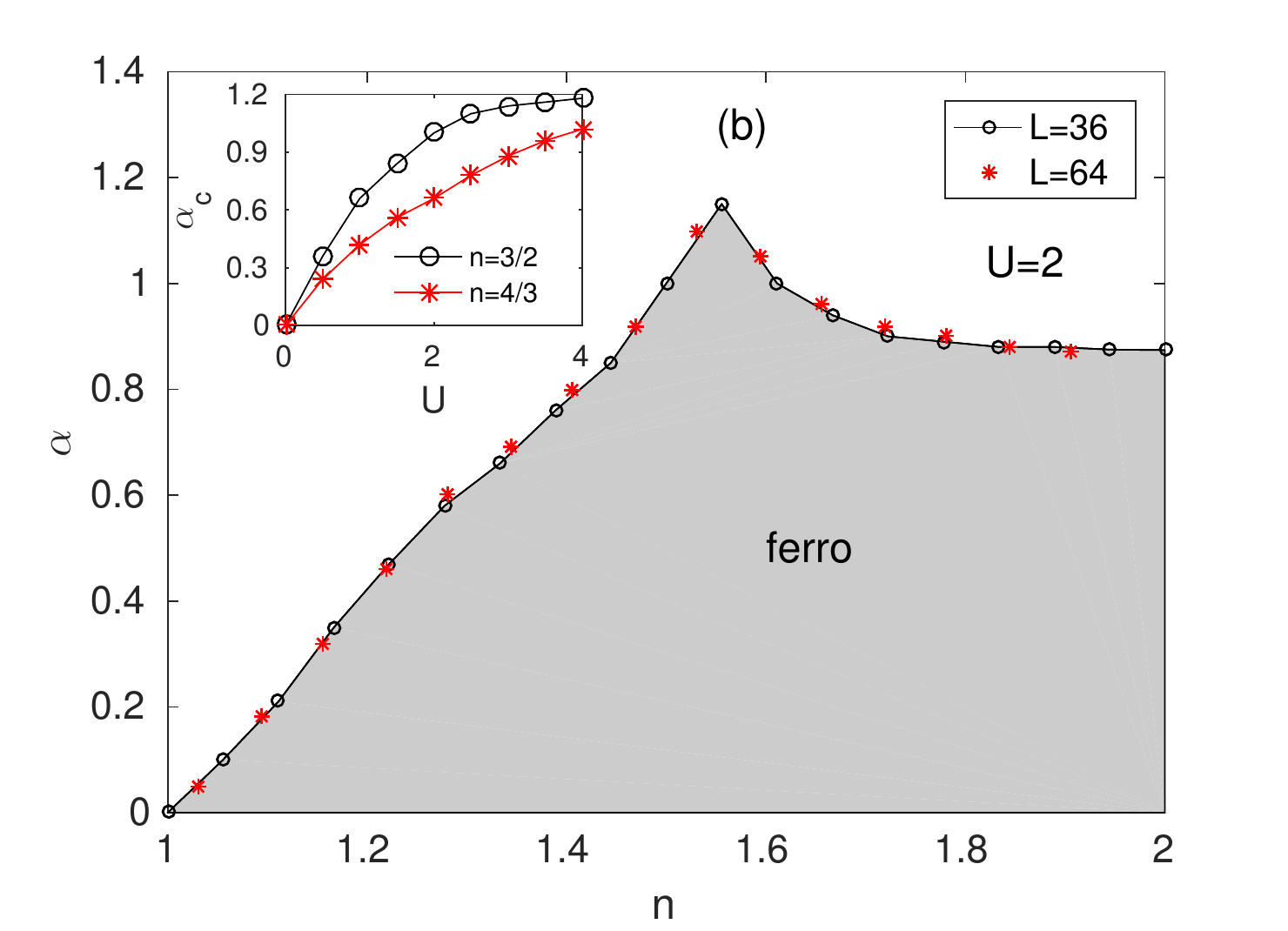}
\end{center}
\caption{(Colour online) (a) The difference $\Delta E=E_f-E_{\text{min}}$ between the  
ferromagnetic state $E_f$ and the lowest ground-state energy       
from $E_g(S_z)$ as a function of the long-range hopping parameter 
$\alpha$ calculated for three different electron concentrations on the 
finite cluster of $L=6 \times 6$ sites ($U=2$). The inset shows $\Delta E$
calculated for $n=3/2$ on clusters of $L=6 \times 6$ and $L=8 \times 8$
sites. (b) The ground-state phase diagram of the generalized Hubbard model
with long-range hopping on the SSL calculated for two different finite
clusters of $L=36$ and $L=64$ sites at $U=2$. The inset shows the critical
value of the long-range hopping parameter $\alpha_c$ (bellow which the
ground state is ferromagnetic) as a function of the Coulomb interaction $U$
calculated for two different electron concentrations $n=3/2$ and $n=4/3$
at $U=2$ and $L=36$~\cite{Fark7}.}
\label{fig18}
\end{figure}

To verify these conjectures, we  performed  exhaustive numerical
studies of the model for a wide range of the model parameters $U,\alpha$
for all even electron concentrations above the half-filled band case.
The nature of the ground state of the Hubbard model 
on the SSL is identified by the projector quantum Monte Carlo 
method~\cite{Imada} with $\theta \sim 30$ and a time slice of $\Delta \theta = 0.05$ 
which suffices to reach well converged values of the observables. 
Typical results of our PQMC calculations obtained on 
the finite cluster of $L=6\times6$ sites for three different electron
concentrations ($n=7/6, 4/3$ and $n=3/2$) are shown in figure~\ref{fig18}a.
It is seen that
for all electron concentrations, there exists a finite  interval of $\alpha$
values where $\Delta E=0$ indicating the fact that ferromagnetism in the 
generalized Hubbard model with exponentially decaying hopping amplitudes 
is not restricted only to high electron concentrations, as in the case of 
the Hubbard model on the SSL with the first, second and third nearest-neighbor 
electron hopping,  but also  extends  to smaller values
of electron concentrations. To exclude the finite-size effect, we  also
performed the same calculations on the larger cluster of $L=8 \times 8$
sites. One can see that the finite-size effects on the stability
region of the ferromagnetic phase are negligible and thus these
results can be satisfactorily extrapolated to the thermodynamic limit
$L=\to \infty$.     

The second step of our study was to specify more precisely the stability 
region of the ferromagnetic state. For this reason, we  performed  
exhaustive numerical studies of the model for a wide range of model parameters
$U, \alpha$ and $n$. The results obtained are presented in the form
of the ground-state phase diagram in the $n-\alpha$ plane (see
figure~\ref{fig18}b).
These results clearly support our aforementioned conjecture that 
ferromagnetic state in the Hubbard model with the exponentially decaying
hopping amplitudes on the SSL is robust and exists for all electron
concentrations above the half-filled band case $n > 1$. The width of the
ferromagnetic domain very strongly depends  on the values of electron
concentrations and reaches its maximum for intermediate values of $n$.
Again, we  performed the evaluation of the finite size effects by 
calculating the same phase diagram on the larger $8\times 8$ cluster, but no
significant effects were observed and thus these results can be also satisfactorily
extrapolated  to the thermodynamic limit.

 We also performed  the same calculations for different values
of the Coulomb interaction in order to demonstrate the interplay between
the long-range electron hopping and the on-site electron interaction. 
The results of our numerical calculations obtained for two different
electron concentrations ($n=4/3$ and $n=3/2$) are summarized in 
the inset to figure~\ref{fig18}a  and they clearly 
demonstrate strong effects
of the Coulomb interaction that  considerably shifts $\alpha_c$ 
to higher values. This leads to a very important conclusion, 
 namely, that correlation effects can, under some conditions 
(e.g., a special lattice structure), stabilize the ferromagnetic 
state in strongly correlated systems.

To reveal the physical limits of the model,  in table~1 we  presented
the actual values of the first, second, third, fourth and fifth
nearest-neighbour hopping amplitudes on the real Archimedean lattice,
corresponding to the real structure of rare-earth tetraborides,
for selected values of the model parameter~$\alpha$. 
\begin{table}[!t]
\caption{ The hopping amplitudes $t'_i$ for the first, second, third,
forth and fifth nearest neighbors on the real Archimedean lattice. Note that
on the real Archimedean lattice corresponding to rare-earth Shastry-Sutherland 
magnets $t'_1=t_1=t_2$~\cite{Fark7}.}
\vspace{3mm}
\begin{center}   
\begin{tabular}{|c|c|c|c|c|c|}
\hline\hline
$\alpha$ & $t'_1$ & $t'_2$ & $t'_3$ & $t'_4$ & $t'_5$ \\
\hline
   $0.1$ & $1$ & $0.9594 $ & $0.9294 $ & $0.9110 $ & $0.8699 $ \\
   $0.2$ & $1$ & $0.9205 $ & $0.8638 $ & $0.8299 $ & $0.7567 $ \\
   $0.3$ & $1$ & $0.8831 $ & $0.8028 $ & $0.7561 $ & $0.6582 $ \\
   $0.4$ & $1$ & $0.8473 $ & $0.7461 $ & $0.6888 $ & $0.5725 $ \\
   $0.5$ & $1$ & $0.8129 $ & $0.6935 $ & $0.6276 $ & $0.4980 $ \\
   $0.6$ & $1$ & $0.7799 $ & $0.6445 $ & $0.5717 $ & $0.4332 $ \\
   $0.7$ & $1$ & $0.7483 $ & $0.5990 $ & $0.5208 $ & $0.3768 $ \\
   $0.8$ & $1$ & $0.7179 $ & $0.5567 $ & $0.4745 $ & $0.3278 $ \\
   $0.9$ & $1$ & $0.6888 $ & $0.5174 $ & $0.4323 $ & $0.2851 $ \\
   $1.0$ & $1$ & $0.6609 $ & $0.4809 $ & $0.3938 $ & $0.2480 $ \\
\hline\hline
\end{tabular}
\end{center} 
\label{tab6}
\end{table}
One can see that the smallest values of $\alpha$ ($\alpha \leqslant 0.2$) are of interest 
only from the academic point of view (a very small decay of hopping amplitudes
$t'_i$) and of physical interest are only those values of $\alpha$, which are 
greater than $0.2$, representing the realistic situation in the Shastry-Sutherland 
materials. The ferromagnetic domain in this limit is still robust
indicating a significant impact of lattice structure and long-range
hopping on the stabilization of ferromagnetism in the strongly
correlated electron systems.

\section{Conclusion}
In this review we  presented the results of our numerical 
calculations concerning the problem of stabilization of ferromagnetism 
in the generalized Hubbard model,  considered herein as a generic model
for a description of itinerant ferromagnetism in narrow-band systems. 
In particular, we  examined the effects of (i) the long-range hopping, 
(ii) the correlated hopping, (iii) the long-range Coulomb interaction, 
(iv) the flat bands and (v) the lattice structure. We  found that
each of the above mentioned terms plays a significant role in stabilizing
 the ferromagnetic state, and for each of these terms we  determined the
domains in the parametric space of the model where the the ferromagnetic
state is the ground state of the generalized Hubbard model. Our results
can be summarized as follows: (i) It is found that the long-range hopping 
with exponentially decaying hopping amplitudes stabilizes the ferromagnetic 
state for a wide range of electron interactions $U$ and electron concentrations 
$n$ for both the one-dimensional and two-dimensional case. In the one 
dimensional case, the ferromagnetic state is stable (above some critical
value of Coulomb interaction) for all electron concentrations $n>1$, while 
in the two-dimensional case, it is stable only for electron concentrations
from the interval $1<n<n_c<2$, where the critical value of electron
concentration $n_c$  strongly depends on the strength of Coulomb interaction
$U$. (ii) Examining the combined effects of long-range and correlated
hopping, we found that ferromagnetic state  for nonzero~$q,U$ and $n$ is further stabilized with an increasing strength of the correlated
hopping term $t'$. The effect is especially strong for intermediate and strong 
values of $q$. There even exists some critical value of $q$ above which 
the ground state is ferromagnetic for all nonzero $U$. With an increasing $t'$, 
this critical value shifts to lower values of $q$ (that represent a much more 
realistic type of electron hopping) and the ferromagnetic domain 
correspondingly increases. (iii) Similarly, examining the combined
effects of the long-rage hopping and long-range Coulomb interaction (both
considered with exponentially decaying amplitudes), we found
that the long-range interaction plays a crucial role in the
stabilization of the ferromagnetic state for electron concentrations
$n \leqslant 1$, while the long-range hopping for $n > 1$. (iv) With respect
to the influence of flat bands on the formation and stabilization of the
ferromagnetic state within the Hubbard model, we found 
that at fixed $U$, the ferromagnetic state is stabilized
with increasing concentration of holes ($1-n$) in the system, and 
at a fixed $n$, the ferromagnetic state is generally stabilized with an
increasing $U$. (v) The study of the Hubbard model on the SSL 
with the first, second and third couplings showed that there
are strong correlations between ferromagnetism and the shape 
of the noninteracting density of states (the lattice
structure). In particular, it is found that ferromagnetism is stabilized 
only for these values of frustration parameters ($t_1,t_2,t_3$), which 
lead to the single peaked noninterating density of states at the band edge. 
From the moment that two or more peaks appear in the noninteracting density of states 
at the band edge, the ferromagnetic state is suppressed. (vi) In addition,
we  found that the ferromagnetic domain for the case of the first,
second and third nearest neighbors is considerably enhanced, when
long-range hopping with exponentially decaying amplitudes is considered. 
All these results point to the fact that the absence of ferromagnetism 
in the ordinary Hubbard model with the nearest-neighbour hopping and 
on-site Coulomb interaction is obviously the consequence of oversimplified 
description of electron hopping and electron interactions on the lattice.

\section*{Acknowledgements}
 This work was supported by the Slovak Research and Development
Agency under the contracts no. APVV-20-0293, APVV-17-0020, the Slovak Grant 
Agency Vega under the contract No. 2-0112-18 and the projects ITMS 
26230120002, ITMS 26210120002 (Slovak infrastructure for high-performance 
computing) supported by the Research and Development Operational Programme 
funded by the ERDF.

\lastpage
\newpage
\ukrainianpart

\title[Зонний феромагнетизм у вузькозонних металах]
{Зонний феромагнетизм у вузькозонних металах}

\author[П. Фаркашовськи]{П. Фаркашовськи}
\address{
	Інститут експериментальної фізики, Словацька академія наук\\
	вул. Ватсонова 47, 043 53 Кошіце, Словаччина}


\makeukrtitle
	
	\begin{abstract}
	З моменту запровадження у 1963 р., модель Хаббарда стала однією з найбільш популярних в літературі, що використовуються для вивчення колективних явищ у вузькозонних металах (феромагнетизм, переходи ``метал-діелектрик'',
	хвилі зарядової густини, високотемпературна надпровідність).
	Серед усіх цих колективних явищ проблема
	зонного феромагнетизму в моделі Хаббарда має найдовшу історію.
	Незважаючи на вражаючу дослідницьку активність у минулому, розуміння фізиками мікроскопічних механізмів, що призводять до стабілізації зонного феромагнетизму в моделі Хаббарда (вузькозонні метали), поки що є далеко не повним. У цьому огляді представлено наші числові результати з цієї проблеми, отримані точною діагоналізацією для малих кластерів, методом ренормгрупи для матриці густини та квантовим методом Монте-Карло в рамках різних узагальнень моделі Хаббарда. Особливу увагу приділено опису вирішальних механізмів (взаємодій), що сприяють стабілізації феромагнітного стану, а саме: (i) дальніх перескоків, (ii) скорельованих перескоків, (iii) далекосяжної кулонівської взаємодії,
	(iv) плоских зон, (v) структури ґратки. Хоча більшість представлених результатів отримано для одновимірного випадку, але також обговорюється і вплив збільшення розмірності системи на її феромагнітний стан.  
		
	\keywords зонний феромагнетизм, системи скорельованих електронів, модель Габбарда
\end{abstract}

\end{document}